\documentclass[12pt]{article}
\usepackage[dvips]{graphicx} 
\usepackage{amsmath}
\usepackage{tabularx} 
\usepackage{setspace}   
\numberwithin{equation}{section}
   
\newcommand{\ba}{\begin{align}} 
\newcommand{\ea}{\end{align}}

\usepackage{amsfonts} 
\usepackage{amssymb}

\newtheorem{thm}{Theorem}
\newtheorem{cor}[thm]{Corollary}
\newtheorem{lemma}[thm]{Lemma}

\newtheorem{prop}[thm]{Proposition}
\newtheorem{defn}[thm]{Definition}

\def\tr{\hbox{Tr}}

\def\be{\begin{eqnarray}}
\def\ee{\end{eqnarray}}
\def\bee{\begin{eqnarray*}}
\def\eee{\end{eqnarray*}}
\def\bmx{\begin{pmatrix}}
\def\emx{\end{pmatrix}}

\def\ts{\textstyle}
\def\bra{\langle}
\def\ket{\rangle}
\def\kb{ \ket \bra }

\def\rt2{\ts \frac{1}{\sqrt{2}} }

\def\ot{\otimes}

\def\tu{{\rm Tube}}
\def\ln{\log}

\title{Comments on Hastings' Additivity Counterexamples}

\author{Motohisa Fukuda\\
{\small Department of Mathematics}\\
{\small University of California, Davis}
\and Christopher King\\
{\small      Department of Mathematics}\\
{\small Northeastern University,  Boston MA 02115} \\
\and David Moser \\
{\small      Department of Physics}\\
{\small Northeastern University,  Boston MA 02115} }

\begin{document}

\maketitle

\begin{abstract}
Hastings \cite{Hastings08} recently provided a proof of the existence of channels
which violate the additivity conjecture for minimal output entropy. In this paper we
present an expanded version of Hastings' proof. 
In addition to a careful elucidation of the details of the proof, we also present bounds for
the minimal dimensions needed to obtain a counterexample.
\end{abstract}

\tableofcontents

\section{The additivity conjectures}
The classical capacity of a quantum channel is the maximum rate at which classical
information can be reliably transmitted through the channel.
This maximum rate is approached asymptotically with multiple channel uses by
encoding
the classical information in quantum states which
can be reliably distinguished by measurements at the output. In general, in order
to achieve optimal performance, it is necessary to use measurements
which are entangled across the multiple channel outputs. However it was conjectured that
product input states are sufficient to achieve the maximal rate of transmission, in other
words that there is no benefit in using entangled states to encode the classical
information.
This conjecture is closely related to other additivity conjectures of quantum information theory,
as will be explained below.
Recently Hastings  \cite{Hastings08} disproved all of these additivity conjectures by proving the existence of channels
which violate the additivity of minimal output entropy. The purpose of this paper is to present in detail the findings of
Hastings' paper, and also to find bounds for the minimal dimensions needed for this type of counterexample.

We begin by formulating the various additivity conjectures.
The Holevo capacity of a quantum channel $\Phi$ is defined by
\be\label{def:chi}
{\chi}^*(\Phi) = \sup_{\{p_i, \, \rho_i\}} S\Big(\Phi\big(\sum_i p_i \rho_i\big)\Big)
- \sum_i p_i \, S\Big(\Phi(\rho_i)\Big)
\ee
where the supremum runs over ensembles of input states, and where $S(\rho)$ denotes the
von Neumann entropy of the state $\rho$:
\be\label{def:entropy}
S(\rho) = - \tr \rho \, \log \rho
\ee
The classical information
capacity $C(\Phi)$ is known \cite{Hol98,SW97} to equal the following limit:
\be
C(\Phi) = \lim_{n \rightarrow \infty} \, \frac{1}{n} \, {\chi}^*(\Phi^{\ot n})
\ee
It has been a longstanding conjecture that the classical information capacity 
is in fact equal to the Holevo capacity:
\be\label{Conj1}
\fbox{\rm Conjecture 1} & \qquad & C(\Phi) = {\chi}^*(\Phi)
\ee
Conjecture 1 would be implied by additivity of ${\chi}^*$ over tensor products.
This led to the following conjecture: for all channels $\Phi$ and $\Omega$,
\be\label{conj2}
\fbox{\rm Conjecture 2} \qquad {\chi}^*(\Phi \ot \Omega) =
{\chi}^*(\Phi) + {\chi}^*(\Omega)
\ee
Subsequently a third conjecture appeared, namely the additivity of minimum output entropy:
\be\label{conj3}
\fbox{\rm Conjecture 3} \qquad S_{\min}(\Phi \ot \Omega) =
S_{\min}(\Phi) + S_{\min}(\Omega)
\ee
where $S_{\min}$ is defined by
\be\label{def:Smin}
S_{\min}(\Phi)  = \inf_{\rho} \, S(\Phi(\rho))
\ee
Finally Amosov, Holevo and Werner \cite{AHW00} proposed a generalization of Conjecture 4 with
von Neumann entropy replaced by the Renyi entropy: for all $p \ge 1$
\be\label{conj4}
\fbox{\rm Conjecture 4} \qquad S_{p,\min}(\Phi \ot \Omega) =
S_{p,\min}(\Phi) + S_{p,\min}(\Omega)
\ee
where $S_{p,\min}$ is the minimal Renyi entropy defined for $p \neq 1$ by
\be\label{def:Spmin}
S_{p,\min}(\Phi)  = \inf_{\rho} \, S_p(\Phi(\rho)), \qquad
S_p(\tau) = \frac{1}{1-p} \, \log \tr \, \tau^p
\ee

In 2004 Shor \cite{Shor04} proved the equivalence of several additivity conjectures,
including Conjectures 2 and 3 above. In subsequent work \cite{FukudaWolf07} it was shown that
Conjectures 1 and 2 are equivalent.
The conjectures have been proved in several special cases
\cite{Amos06, Amos07, BFMP00, FujiHashi02, King02, King03, KingNathRus05, KingRuskai01},
but recently most progress has been made in the search for counterexamples. This started
with the Holevo-Werner channel \cite{HolWer02} which provided a counterexample
to Conjecture 4 with $p > 4.79$, then more recently Winter and Hayden found counterexamples
to Conjecture 4 for all $p > 1$ \cite{HaydenWinter08}, and violations have since been
found also for $p=0$ and $p$ close to zero \cite{CHLMW08}. Finally in 2008,
Hastings \cite{Hastings08} produced a family of channels which violate Conjecture 3, namely
additivity of minimal output von Neumann entropy, thereby also proving (via \cite{Shor04} and \cite{FukudaWolf07})
that Conjectures 1 and 2 are false.

The product channels introduced by Hastings have the form
$\Phi \ot \overline{\Phi}$ where $\Phi$ is a special channel which we call
a {\em random unitary channel}. This means that
there are positive numbers $w_1,\dots,w_d$ with $\sum_i w_i = 1$ and unitary $n \times n$ matrices $U_1,\dots,U_d$
such that
\be\label{rand.unit1}
\Phi(\rho) = \sum_{i=1}^d w_i \, U_i \,\rho \,U_i^*, \quad\quad
\overline{\Phi}(\rho) = \sum_{i=1}^d w_i \, \overline{U_i}\, \rho \,\overline{U_i^*}
\ee
These channels are chosen randomly using a distribution that depends on the two integers 
$n$ and $d$, where $n$ is the dimension of the
input space and $d$ is the dimension of the environment. Hastings' main result is that
for $n$ and $d$ sufficiently large there are random unitary channels which violate Conjecture 3, that is
\be\label{Hast1}
S_{\min}(\Phi \ot  \overline{\Phi}) <
S_{\min}(\Phi) + S_{\min}(\overline{\Phi})
\ee

This result also allows a direct construction of channels which violate Conjectures 1 and 2,
as we now show.
Using results from the paper \cite{FukudaWolf07}, the inequality (\ref{Hast1}) implies that
the additivity of minimal output entropy does not hold for the product
$ \Phi^\prime \otimes \Phi^\prime$, where $\Phi^\prime = \Phi \oplus \overline{\Phi}$.
In addition, as shown in the paper \cite{Fuk06}, there is a unital extension of $\Phi^\prime$,
denoted $\Phi^{\prime\prime}$,
such that the additivity of minimal output entropy does not hold for
$ \Phi^{\prime\prime}\otimes\Phi^{\prime\prime}$, and such that
\be
S_{\min} (\Phi^{\prime\prime}\otimes\Phi^{\prime\prime})
= 2\ln D - \chi^* (\Phi^{\prime\prime}\otimes\Phi^{\prime\prime}).
\ee
where $D$ is the dimension of the output space for $\Phi^{\prime\prime}$.
Thus $\Phi^{\prime\prime}$ provides a counterexample for Conjecture 2, and
\be
\lim_{k\rightarrow \infty} \frac{1}{2k}\chi^* ((\Phi^{\prime\prime})^{\otimes 2k}) >
\chi^*(\Phi^{\prime\prime}).
\ee
Therefore, the classical capacity of $\Phi^{\prime\prime}$
does not equal its Holevo capacity, and this provides a counterexample for Conjecture 1.

\medskip
One key ingredient in the proof is the relative sizes of dimensions, namely $n >> d >> 1$,
where $n$ is the dimension of the input space, and $d$ is the dimension of the environment.
Recall that in the Stinespring representation a channel is viewed as a partial isometry
from the input space ${\cal H}_{in}$ to the product of output and environment spaces 
${\cal H}_{out} \ot {\cal H}_{env}$, followed by a partial trace over the environment. 
The image of ${\cal H}_{in}$ under the partial isometry is a subspace of dimension $n$
sitting in the product ${\cal H}_{out} \ot {\cal H}_{env}$.
Making the environment dimension $d$ much smaller than the input dimension $n$ should guarantee
that with high probability this subspace will consist of almost maximally entangled states.
For such states the output entropy will be close to the maximal possible value $\log d$,
and therefore the minimal entropy of the channel should also (hopefully) be close to $\log d$.
At the same time the product channel $\Phi \ot \overline{\Phi}$ sends the maximally entangled state
into an output with one relatively large eigenvalue, and thus one might hope to find a gap between
$S_{\min}(\Phi \ot  \overline{\Phi})$ and $S_{\min}(\Phi) + S_{\min}(\overline{\Phi})$.
Turning this vague notion into a proof requires considerable insight and ingenuity. In this paper we
focus on the technical aspects of Hastings' proof. Some of the estimates and inequalities
derived in this paper are new, but all the main ideas and methods are taken from
\cite{Hastings08}.

\medskip
The paper is organized as follows. In Section 2 we define notation and make a precise statement
of Hastings' results. In Section 3 we present some background material on probability
distributions for states and channels. In Section 4 we `walk through' the proof of Hastings' Theorem,
stating results where needed and delineating the logic of the argument.
In Section 5 we give the proofs of various results needed in Section 4 and elsewhere.
Section 6 discusses different aspects of the proof and possible directions for further research.
The Appendix contains the derivation of some estimates needed for the proof.

\section{Notation and statement of results}
We will mostly avoid Dirac bra and ket notation, although it will be used
in Sections \ref{pf:lemma01} and \ref{pf:lem:uniform-lwr-bnd}.

\subsection{Notation}
Let ${\cal M}_n$ denote the algebra of complex $n \times n$ matrices.
The identity matrix will be denoted $I_n$, or just $I$.
The set of {\em states} in ${\cal M}_n$ is defined as
\be
{\cal S}_n = \{ \rho \in {\cal M}_n \,:\, \rho = \rho^* \ge 0, \,\, \tr \rho  =1 \}
\ee

\medskip
\noindent The set of {\em unit vectors} in $\mathbb{C}^n$ will be denoted
\be
{\cal V}_n = \{ z  = (z_1,\dots,z_n)^T \in \mathbb{C}^n \,:\, z^* z = \sum_{i=1}^n |z_i|^2 = 1 \}
\ee

\medskip
Every unit vector $ z  \in {\cal V}_n$ defines a pure state $\rho = z z^*$
satisfying $\rho^2 = \rho$.
The set of unit vectors ${\cal V}_n$ is identified with the real $(2n -1)$-dimensional sphere $S^{2n-1}$, and hence carries
a unique uniform probability measure which we denote $\sigma_n$. 

\medskip
\noindent The set of unitary matrices in ${\cal M}_n$ is denoted
\be
{\cal U}(n) = \{ U \in {\cal M}_n \,:\, U U^* = I \}
\ee
We will write $H_n$ for the normalized Haar measure on ${\cal U}(n)$.

\medskip
A {\em channel} is a completely positive trace-preserving map
$\Phi \,:\, {\cal M}_n \rightarrow {\cal M}_m$. 
Recall the definition of random unitary channel (\ref{rand.unit1}):
\be
\Phi(\rho) = \sum_{i=1}^d w_i \, U_i \,\rho \,U_i^*
\ee
The set of all random unitary channels on ${\cal M}_n$ with $d$ summands will be denoted ${\cal R}_d(n)$.
Given a channel $\Phi \in {\cal R}_d(n)$
the {\em complementary} or {\em conjugate} channel $\Phi^C : \, {\cal M}_n \rightarrow {\cal M}_d$ is defined by
\cite{Holevo05}, \cite{KMNR}
\be\label{def:Phi-C}
\Phi^C(\rho) = \sum_{i,j =1}^d \, \sqrt{w_i w_j} \,\, \tr (\rho \, U_j^* U_i) \,\, | i \kb j |
\ee
As is well-known, for any input state $\rho$ the output states 
$\Phi(\rho)$ and $\Phi^C(\rho)$ are related by
\be\label{Phi&PhiC}
\Phi(\rho) = \tr_2 \,W \rho W^*, \quad
\Phi^C(\rho) = \tr_1 \,W \rho  W^*
\ee
Here,  $W:\mathbb{C}^n \rightarrow \mathbb{C}^{nd}$ is a partial isometry. 
Also $\tr_2$ denotes the partial trace over the state space of the environment,
and $\tr_1$ denotes the partial trace over the state space of the system.
When $\rho = z z^*$ is a pure state, the matrices $\Phi(z z^*)$ and $\Phi^C(z z^*)$ are
partial traces of the same pure state, and thus have the same non-zero spectrum and the same
entropy. Therefore $S_{\min}(\Phi) = S_{\min}(\Phi^C)$. For the purposes of constructing the
counterexample it is convenient to work with both $\Phi$ and $\Phi^C$. 
In particular, we are interested in the cases where $W$ consists of
rescaled unitary block matrices;
\be
W = \begin{pmatrix}
\sqrt{w_1} U_1 \\ \vdots \\ \sqrt{w_d}U_d,
\end{pmatrix}
\ee
Note that $\sum_iw_i =1$ as $W$ is a partial isometry.
We define a measure on this subset of partial isometries, in Section \ref{ProbChannel},
as the product of Haar measures and a particular measure on the simplex.

\medskip
The complex conjugate channel $\overline{\Phi}$ is defined by
\be
\overline{\Phi}(\rho) = \sum_{i=1}^d w_i \, \overline{U_i} \rho \overline{U_i^*}
=  \sum_{i=1}^d w_i \, \overline{U_i} \rho U_i^T
\ee
Again note that $\Phi$ and $\overline{\Phi}$ have identical minimum output entropies.

\subsection{The main result}
Following the work of Winter and Hayden \cite{HaydenWinter08}, the counterexample is taken to be a product channel of the form
$\Phi \ot \overline{\Phi}$ where $\Phi$ is a random unitary channel.
Hastings first proves the following
universal upper bound for the minimum output entropy of such a product.

\begin{lemma}\label{lemma01}
For any $\Phi \in  {\cal R}_d(n)$,
\be
S_{\min}(\Phi \ot \overline{\Phi}) \le 2 \log d - \frac{\log d}{d}
\ee
\end{lemma}

Lemma \ref{lemma01} will be proved in Section \ref{pf:lemma01}. The counterexample is found by proving the existence of a 
random unitary channel $\Phi$ whose minimum output entropy
is greater than one half of this upper bound, that is greater than $\log d - \log d /2d$. For such a channel it will follow that
\be
S_{\min}(\Phi \ot \overline{\Phi}) &\le & 2 \log d - \frac{\log d}{d} \\
& < & 2 S_{\min}(\Phi) \\
& = & S_{\min}(\Phi) + S_{\min}(\overline{\Phi})
\ee
and this will provide the counterexample to Conjecture 3.
Hastings \cite{Hastings08} proved the existence of such channels using a combination
of probabilistic arguments and estimates involving the distribution of the reduced
density matrix of a random pure state. The next Theorem is a precise statement of
Hastings' result.

\begin{thm}\label{thm2}
There is $h_{\min} < \infty$, such that for all $h > h_{\min}$,  all
$d$ satisfying $d \log d \ge h$, and all $n$ sufficiently large,
there is $\Phi \in {\cal R}_d(n)$ satisfying
\be\label{main.lower}
S_{\min}(\Phi) > \log d - \frac{h}{d}
\ee
\end{thm}

By taking $d$ large enough so that $2 h_{\min} < \log d$, we deduce from Theorem \ref{thm2}
that there is a channel $\Phi$ satisfying
\be
S_{\min}(\Phi) > \log d - \frac{\log d}{2 d}
\ee
and this establishes the existence of counterexamples for Conjecture 3.
In fact the proof will show that as $d, \, n \rightarrow \infty$, the probability that a randomly chosen channel in
${\cal R}_d(n)$ will satisfy the bound (\ref{main.lower}) approaches one. 

\medskip
It would be interesting to determine
the set of integers $(n,d)$ for which there are random unitary channels in ${\cal R}_d(n)$ violating additivity,
and in particular to find the smallest dimensions which allow violations, as well as the size of the largest possible violation.
Following this line of reasoning we define
\be\label{def:crit-dims}
d_{\min} & = & \inf \Big\{d \,:\, \exists \, n, \, \exists \, \Phi \in {\cal R}_d(n) \,{\rm s.t.} \, 
S_{\min}(\Phi) > \log d - \frac{\log d}{2 d} \Big\} \nonumber \\
n_{\min} & = & 
\inf \Big\{n \,:\, \exists \, d, \, \exists \, \Phi \in {\cal R}_d(n) \,{\rm s.t.} \, 
S_{\min}(\Phi) > \log d - \frac{\log d}{2 d} \Big\} \nonumber \\
\Delta S_{\max} & = & \sup_{n,d} \sup_{\Phi \in {\cal R}_d(n)}
\Big(S_{\min}(\Phi) + S_{\min}(\overline{\Phi}) - S_{\min}(\Phi \ot \overline{\Phi})\Big)
\ee

The next result gives some bounds on these quantities.

\begin{prop}\label{prop:min-dims}
\bee
d_{\min} & < &  3.9 \times 10^{4} \nonumber \\
n_{\min} & < & 7.8 \times 10^{32} \nonumber \\
\Delta S_{\max} & > & 9.5 \times 10^{-6}
\eee
\end{prop}

Proposition \ref{prop:min-dims} will be proved in Section \ref{pf:prop:min-dims}.
The bounds in Proposition \ref{prop:min-dims} are surely not optimal, however they may
indicate the delicacy of the non-additivity effect for this class of channels.
It would certainly be interesting to tune the estimates in this paper in order to improve the
bounds in Proposition \ref{prop:min-dims},
or even better to find a different class of channels where the effect is larger.

\section{Background on random states and channels}
As mentioned above, the proof of Theorem \ref{thm2} relies on probabilistic arguments,
involving distributions of pure states and random unitary channels. The next sections explain 
the distributions which play a role in the proof.

\subsection{Probability distributions for states}
Recall that ${\cal V}_n$ is the set of unit vectors in $\mathbb{C}^n$. This set carries a natural
uniform measure $\sigma_n$, namely the uniform measure on the (real) $(2n-1)$-dimensional
sphere.
If $\mathbb{C}^{d n} = \mathbb{C}^d \ot \mathbb{C}^n$ is a product space, then
a unit vector $z \in {\cal V}_{d n}$ can be written as a $n \times d$ matrix $M$, with entries
\be
M_{ij}(z) = z_{(i-1)d + j}, \quad i=1,\dots n, \,\, j=1,\dots,d
\ee
satisfying $\tr M^* M = \sum_{ij} |z_{ij}|^2 = 1$.
Define the map $G \,:\, {\cal V}_{d n} \rightarrow {\cal M}_d$ by
\be\label{def:G}
G(z) = M(z)^* M(z)
\ee
It follows that $G(z) \ge 0$ and $\tr \, G(z) = 1$, and hence the image of $G$ lies in ${\cal S}_d$
(the set of $d$-dimensional states). Since $z$ is a random vector (with distribution $\sigma_{dn}$) it follows that $G(z)$ is a 
${\cal S}_d$-valued random variable, or more simply a random state.
Its distribution has been studied in many other contexts (see for example \cite{HayLeuWin06}) and it plays
a key role in the proof here.

\subsection{Probability distributions on the simplex $\Delta_d$}\label{subsect:nu,mu}
Let ${\Delta}_d$ denote the simplex of $d$-dimensional probability distributions:
\be\label{def:Delta}
{\Delta}_d = \{(x_1,\dots,x_d) \subset \mathbb{R}^d \,:\, x_i \ge 0, \,\, \sum_{i=1}^d x_i = 1 \}
\ee
We define below three different probability distributions on $\Delta_d$. One is the uniform
measure inherited from $\mathbb{R}^d$, and the others are  defined by the 
diagonal entries and the eigenvalues of $G(z)$ where $z$ is a random unit vector in ${\cal V}_{dn}$.

\medskip
\noindent \fbox{\rm Uniform distribution}
The simplex $\Delta_d$  carries a natural measure inherited from Lebesgue measure on $\mathbb{R}^d$:
this is conveniently written as
\be\label{def:delta}
\delta\bigg(\sum_{i=1}^d w_i - 1\bigg) dw_1 \dots d w_d = \delta\bigg(\sum_{i=1}^d w_i - 1\bigg) \, [dw]
\ee
where $\delta(\cdot)$ is the Dirac $\delta$-function. Integrals with respect to
this measure can be evaluated
by introducing local coordinates on 
$\mathbb{R}^d$ in a neighborhood of $\Delta_d$. 
In particular the volume of $\Delta_d$ with respect to the measure (\ref{def:delta})
can be computed:
\be\label{Delta-vol}
\int_{\Delta_d} \delta\bigg(\sum_{i=1}^d w_i - 1\bigg) \, [dw] = \frac{1}{(d-1)!}
\ee

\medskip
\noindent \fbox{\rm Diagonal distribution $\nu_{d,n}$}
Let $z \in {\cal V}_{dn}$ be a random unit vector in $\mathbb{C}^{n} \ot \mathbb{C}^d$.
The joint distribution of the diagonal entries $(G_{11}(z),\dots,G_{dd}(z))$ will be denoted
$\nu_{d,n}$. It is possible to find an explicit formula for the density of $\nu_{d,n}$, however we will
not need it in this paper.
It is sufficient to note that a collection of $d$ random variables $Y_1,\dots,Y_d$ have the joint
distribution $\nu_{d,n}$ if and only if they can be written as 
\be\label{nu-Y's}
Y_j = \sum_{i=1}^n | z_{ij} |^2, \quad j=1,\dots,d
\ee
where $\{z_{ij}\}$ are 
the components of a uniform random vector on the unit sphere in $\mathbb{C}^{n} \ot \mathbb{C}^d$.
We come back to this problem in Section \ref{pf:lem1}.

\medskip
\noindent \fbox{\rm Eigenvalue distribution $\mu_{d,n}$}
As noted above
the eigenvalues of $G(z)$ are non-negative and sum to one. 
\footnote{
$G(z)$ gives the complex Wishart matrix when 
$z \in \mathbb{C}^{n} \ot \mathbb{C}^d$ with each entry $z_{ij}$ being IID complex normal distribution.
The eigenvalue distribution was shown to be
proportional to $\prod_{1 \le i < j \le d} (w_i - w_j)^2 
\prod_{i=1}^d w_i^{n-d}[dw]$, for example, in \cite{Bronk}.
}
However the eigenvalues are not ordered and 
so define a map not into $\Delta_d$ but rather into the quotient
$\Delta_d/\Sigma_d$ where $\Sigma_d$ is the symmetric group. 
Thus when $z \in {\cal V}_{dn}$ is a random vector the eigenvalues
of $G(z)$ are $\Delta_d/\Sigma_d$-valued random variables.
However it is convenient to use a joint density for the eigenvalue distribution on $\Delta_d$,
with the understanding that it should be evaluated only on events which are invariant under
$\Sigma_d$. This density is known explicitly \cite{LloydPagels}, \cite{ZycSom01}:
for any event $A \subset \Delta_d$
\be\label{mu-density}
\mu_{d,n}(A) = Z(n,d)^{-1} \, \int_{A} \prod_{1 \le i < j \le d} (w_i - w_j)^2 
\prod_{i=1}^d w_i^{n-d} \, \delta\bigg(\sum_{i=1}^d w_i - 1\bigg) \, [dw]
\ee
where $Z(n,d)$ is a normalization factor. 
The distribution $\mu_{d,n}$ plays an essential role in the proof of Theorem \ref{thm2}.
Explicit expressions for $Z(n,d)$ are known \cite{ZycSom01}.
In Appendix A we derive the following bound:
for $n$ sufficiently large,
\be\label{Z-est1}
Z(n,d)^{-1} \le n^{d^2} \, d^{d \,(n-d)}
\ee

\subsection{Estimates for $\mu_{d,n}$}
Define the function
\be\label{def:F}
\fbox{$F(x) = - \log x + x - 1$}
\ee

\begin{lemma}\label{lem:bound-mu}
For all $d$, for $n$ sufficiently large, and for any event $A \subset {\Delta}_d$,
\be\label{mu2}
\mu_{d,n}(A) \le \int_A \exp \Big[d^2 \log n - (n-d) \sum_{i=1}^d \, F(d w_i) \Big]
\, \delta\bigg(\sum_{i=1}^d w_i - 1\bigg) \, [dw]
\ee
\end{lemma}

This Lemma will be proved in Section \ref{pf:lem:bound-mu}. Using (\ref{Delta-vol}) we
immediately get the following bound.

\begin{cor}\label{cor1}
For all $d$, for $n$ sufficiently large, and for any event $A \subset {\Delta}_d$,
\be\label{mu3a}
\mu_{d,n}(A) 
\le  \exp \Big[d^2 \log n - \log (d-1)! - (n-d) \inf_{w \in A} \sum_{i=1}^d F(d w_i) \Big]
\ee
\end{cor}

Note that $F(x)$ is convex, and also $F(1) = F'(1) = 0$.
The Taylor expansion around $1$ gives
\be\label{taylor}
F\left(1 + d \delta w  \right) =  \frac{1}{2}  d^2 (\delta w)^2 + R
\label{exp2}
\ee
where the remainder is
\be\label{remainder}
R = - \frac{1}{3!}\left(1 +  d \delta \right)^{-3} (d \delta w)^3.
\ee
and $\delta $ is some value between $0$ and $\delta w$.
Note that $- 1/d < \delta w < (d-1)/d$ as $0\leq w \leq 1$.
Also, $R>0$ if $\delta w <0$.
When $\delta w >0$
\be
0>R> - \frac{1}{6}d^3(\delta w)^3.
\label{exp3}
\ee

\medskip
Recall that $F(x) \ge 0$, so we have the bound $F(d w_i) \ge 0$ for all $i$.
Thus feeding (\ref{taylor}) into Corollary \ref{cor1} gives the following estimate, which will be used in Section
\ref{pf:lem:prob-Tc}.

\begin{cor}\label{cor2}
For all $d$, for $n$ sufficiently large, and for any $i=1,\dots,d$,
\be\label{mu3}
\mu_{d,n}\bigg\{ w \,:\, \Big|w_i - \frac{1}{d}\Big| \ge t \bigg\} 
\le  \exp \Big[d^2 \log n - \log (d-1)! - \frac{n-d}{2} d^2 t^2 + \frac{n-d}{6} d^3 t^3 \Big]
\ee
\end{cor}

\subsection{Probability distribution for random unitary channels} \label{ProbChannel}
A random unitary channel (\ref{rand.unit1}) is determined by the coefficients $w_i$ and the unitary
matrices $U_i$. Thus
the set of random unitary channels ${\cal R}_d(n)$ is naturally identified with  $\Delta_d \times {\cal U}(n)^d$. 
Recall the distribution $\nu_{d,n}$ defined in Section \ref{subsect:nu,mu} for the diagonal entries of $G(z)$,
and the Haar measure $H_n$ defined on ${\cal U}(n)$.
We define the following product probability measure on ${\cal R}_d(n)$:
\be\label{def:P}
{\cal P}_{d,n} = \nu_{d,n} \times H_n \times \cdots \times H_n
\ee
where $H_n \times \cdots \times H_n$ is the $d$-fold product Haar measure on ${\cal U}(n)^d$.
Using  the measure ${\cal P}_{d,n}$ on ${\cal R}_d(n)$ means that the unitaries $U_i$ are selected randomly and independently, while the coefficients $w_j$ 
have the joint distribution $\nu_{d,n}$, and thus can be written in the form (\ref{nu-Y's})
where $\{z_{ij}\}$ ($i=1,\dots,n$; $j=1,\dots,d$) are the components of a random unit vector in
${\cal V}_{nd}$.

\medskip
Recall the definition (\ref{def:Phi-C})  of the conjugate channel.
Define the map
\be\label{def:H}
H: {\cal R}_d(n) \times {\cal V}_n \rightarrow {\cal M}_d, \quad
(\Phi, z) \mapsto \Phi^C(z z^*)
\ee
Recall the definition
(\ref{def:G}) of the map $G \,:\, {\cal V}_{d n} \rightarrow {\cal M}_d$.
The following relation between the distributions ${\cal P}_{d,n}$, $\sigma_n$ and $\sigma_{dn}$ is crucial to the proof. 

\medskip
\begin{lemma}\label{lem1}
\be
H^*({\cal P}_{d,n} \times \sigma_n) = G^*(\sigma_{d n})
\ee
\end{lemma}

Lemma \ref{lem1} will be proved in Section \ref{pf:lem1}.
It implies that if $\Phi$ is chosen randomly according to
the measure ${\cal P}_{d,n}$ and $z$ is chosen randomly and uniformly in ${\cal V}_n$, then
the eigenvalues of the matrix
$\Phi^C(z z^*)$ will have the distribution $\mu_{d,n}$.

\section{Proof of Theorem \ref{thm2}}
The main idea of the proof is to isolate some properties of random unitary channels which
are typical for large values $n$ and $d$. These  properties will  then be
used to prove that large minimum output entropy is also typical for random unitary channels
when $n$, $d$ are large. 

\medskip
Recall that the environment dimension $d$ will be chosen to be much smaller than the input dimension $n$.
As the identity (\ref{Phi&PhiC}) shows,
selecting a channel in ${\cal R}_d(n)$ corresponds to selecting a 
subspace of dimension $n$  in the product space $\mathbb{C}^n \ot \mathbb{C}^d$. The structure of
random bipartite subspaces was analyzed in the paper \cite{HayLeuWin06}, and it was shown that in some
circumstances most states in a randomly selected subspace will be close to maximally entangled.
In such a situation the reduced density matrix
of a randomly selected output state $\Phi^C(z z^*)$ will be close to the maximally mixed state $I/d$,
and hence its entropy will be close to $\log d$. Although this
observation plays an essential role
in Hastings' proof, the
methods used in \cite{HayLeuWin06} do not directly yield  the bounds needed.

\subsection{Definition of the typical channel}
A channel $\Phi$ will be called typical if $\Phi^C$
maps at least one half of input states into a small ball centered at the maximally
mixed output state $I/d$. The size of the small ball in question involves a numerical parameter $b$
and is defined as follows:
\be
B_d(n) = \bigg\{ \rho \in {\cal S}_d \,:\, \Big\| \rho - \frac{1}{d} I \Big\|_{\infty} \le b \sqrt{\frac{\log n}{n}} \bigg\}
\ee

\begin{defn}
A random unitary channel $\Phi$ is called {\em typical} if with probability at least $1/2$ a randomly
chosen input state is mapped by $\Phi^C$ into the set $B_d(n)$. The set of typical channels is denoted $T$:
\be
T = \bigg\{ \Phi \,:\, \sigma_n\Big(z \,:\, \Phi^C(z z^*) \in B_d(n)\Big) \ge 1/2 \bigg\}
\ee
\end{defn}

As the next result shows, for large $n$ most channels are typical.

\begin{lemma}\label{lem:prob-Tc}
For every $b > \sqrt{3}$ there is $\alpha > 0$ such that for $n$ sufficiently large,
and for all $d$
\be\label{lem:prob-Tc1}
{\cal P}_{d,n}(T^c) \le \frac{2 \, d}{(d-1)!} \,\, \exp[-\alpha \, d^2 \, \log n]
\ee
\end{lemma}

Thus if $b > \sqrt{3}$, then as $n \rightarrow \infty$ with high probability a randomly chosen channel will  lie in the set $T$.
In particular ${\cal P}_{d,n}(T^c) < 1$ for $n$ sufficiently large. The number $\alpha$ can be chosen to satisfy
\be\label{def:alpha}
\alpha =  \frac{b^2(n-d)}{3n} - 1
\ee
The dimension $n$ must be large enough so that the right side of (\ref{def:alpha}) is positive, and also
so that  $n \ge b^2 d^2 \, \log n$ (this is a technical condition needed in the proof, see Section \ref{pf:lem:prob-Tc}).

\medskip
The second property of a typical channel $\Phi$ is the existence of a `tube' of output states surrounding
$\Phi^C(z z^*)$  for every input state $z \in {\cal V}_n$.
This property is used to eliminate the possibility of isolated output states with low entropy:
if for some $z$ the output entropy $S(\Phi^C(z z^*))$ is small, then there is a nonzero fraction of
input states whose outputs also have low entropy. In order to define the tube we first construct
a line segment $Y(\rho)$ pointing from a general state $\rho$ toward the maximally mixed state $I/d$.
The length of the segment depends on a parameter $\gamma$, which satisfies
$0 < \gamma < 1$:
\be\label{def:Y}
Y(\rho) = \Big\{ r \rho + (1-r) \frac{1}{d} I \,:\, \gamma \le r \le 1 \Big\}
\ee

\medskip
The tube at $\rho$ is defined to be the set of states which lie within a small distance of the set $Y(\rho)$, and thus form a thickened line segment pointing from $\rho$ toward the maximally mixed state. The definition of `small' here depends on 
the size of the ball $B_d(n)$, and also on another numerical parameter $t$.

\begin{defn}
Let $\rho \in {\cal S}_d$, then the {\em $\tu$ at $\rho$} is defined as
\be\label{def:tube}
\tu(\rho) = \bigg\{ \theta \in {\cal S}_d \,:\, dist(\theta,Y(\rho)) \le t \, \sqrt{\frac{d \log n}{n}} \bigg\},\quad
dist(\theta,Y(\rho))=
\inf_{\tau \in Y(\rho)} \| \theta - \tau \|_{\infty}
\ee
\end{defn}
The next result shows that for a channel $\Phi$ in the typical set $T$, 
and for any state $\rho = \Phi^C(z z^*)$ in the image of $\Phi^C$, there is a uniform lower
bound for the probability that a randomly chosen state belongs to the tube at $\rho$.
As explained before, this means that an output state $ \Phi^C(z z^*)$ cannot be too isolated from the other output states.

\begin{lemma}\label{lem:uniform-lwr-bnd}
For all $d \ge 3$ there is $\beta > 0$ such that for $n$ sufficiently large,
for all $t \ge b+4$, and for all
$\Phi \in T$ and $\rho \in {\rm Im}(\Phi^C)$,
\be
\sigma_n\Big(z \,:\, \Phi^C(z z^*) \in \tu(\rho)\Big) \ge \beta \, \bigg( 1-\gamma \bigg)^{n-1}
\ee
\end{lemma}
Lemma \ref{lem:uniform-lwr-bnd} will be proved in Section \ref{pf:lem:uniform-lwr-bnd}.
The number $\beta$ is given by the following expression:
\be\label{def:beta}
\beta = \frac{1}{2} - (d^2 + 2) \bigg(1 - \frac{d \log d}{n} \bigg)^{n-1}
\ee
It can be easily seen that for all $d \ge 3$ the right side of (\ref{def:beta}) is positive
for $n$ sufficiently large.

\subsection{Definition of the low-entropy events $E$}
Define the set of channels whose minimum output entropy does not satisfy our requirements for a violation:
\be\label{def:C-d,n}
C_{d,n} = \bigg\{ \Phi \in {\cal R}_d(n)\,:\, S_{\min}(\Phi) \le \log d - \frac{h}{d} \bigg\}
\ee
The goal is to show that for $d$, $n$ and $h$ sufficiently large we have ${\cal P}_{d,n}(C_{d,n}) < 1$, implying 
that ${\cal P}_{d,n}(C_{d,n}^c) > 0$, and thus that there exist random unitary
channels with $S_{\min}(\Phi) > \log d - h/d$. The proof will hold for all $h, d$ sufficiently large, and thus by
taking $\log d \ge 2 h$ this will provide a counterexample to
additivity.
The method is to find useful upper and lower bounds for the probability of a particular event $E$
in ${\cal R}_d(n) \times {\cal V}_n$. The event $E$ is chosen to contain all the pairs $(\Phi, z)$ where
$\Phi^C(z z^*)$ lies in a tube connected to a state of low entropy.
This set of tubes is defined by
\be\label{def:J}
J = \bigcup_{\rho} \bigg\{ \tu(\rho)\,:\, S(\rho) \le \log d - \frac{h}{d} \bigg\}
\ee
Then the main event of interest for us is the following subset of
${\cal R}_d(n) \times {\cal V}_n$:
\be
E = \{ (\Phi, z) \,:\, \Phi^C(z z^*) \in J \} = H^{-1}(J)
\ee
where $H$ is the map defined in (\ref{def:H}).
The proof will proceed by proving upper and lower bounds for 
the probability of $E$, that is $({\cal P}_{d,n} \times \sigma_n)(E)$.
These bounds will hold
for any $0 < \gamma < 1$; the
parameter $\gamma$ will be `tuned' at the end in order to derive an estimate for the minimal size $h_{\min}$ needed for the counterexample.
As noted the construction works for any values of the parameters $b$, $t$ satisfying
$b > \sqrt{3}$ and $t \ge b+4$. The sizes of $b$ and $t$ do not play a crucial role, and they can be set
to the values $b=2$ and $t=6$ without changing anything in the proof.

\subsection{The upper bound for $Prob(E)$ }
Note that by Lemma \ref{lem1},
\be\label{uppr1}
({\cal P}_{d,n} \times \sigma_n)(E) & = &({\cal P}_{d,n} \times \sigma_n)(H^{-1}(J)) \nonumber \\
&=& H^*({\cal P}_{d,n} \times \sigma_n)(J) =
G^*(\sigma_{d n})(J)
\ee
Let $\rho$ be a fixed state in the set of tubes $J$. Then by definition there is a state $\tau \in {\cal S}_d$ with low entropy such that
$\rho$ lies in the tube at $\tau$. Thus for some $r$ satisfying $\gamma \le r \le 1$
\be\label{infinity-cond1}
\bigg\| \rho -  \bigg(r \tau + (1-r) \frac{1}{d} I\bigg) \bigg\|_{\infty} \le t \sqrt{\frac{d \log n}{n}}, \quad
S(\tau) \le \log d - \frac{h}{d}
\ee
Letting $q_i, p_i$ denote the eigenvalues of $\rho$, $\tau$ 
respectively, it follows that
\be\label{def:eps1}
q_i = r p_i + (1-r) \frac{1}{d} + \epsilon_i, \quad i=1,\dots,d
\ee
where $p_i$, $\epsilon_i$ satisfy
\be\label{ent.cond1}
- \sum_i p_i \log p_i \le \log d - \frac{h}{d}, \qquad
\sum_{i=1}^d \epsilon_i = 0
\ee
Weyl's inequality and (\ref{infinity-cond1}) imply that
\be\label{eps.cond1}
| \epsilon_i | \le t \sqrt{\frac{d \log n}{n}} 
\ee
The entropy condition (\ref{ent.cond1}) can be written as
\be\label{ent.cond2}
\sum_{i=1}^d p_i d \, \log (p_i d) = 
\sum_{i=1}^d \bigg(p_i d \, \log (p_i d) - p_i d + 1 \bigg) \ge h
\ee
Define the function
\be\label{def:f}
\fbox{$f(x) = x \, \log x - x + 1$}
\ee

\begin{lemma}\label{lem:f-bound}
\be\label{lem:f-bound1}
\sup_{x \ge 0, \,\, \gamma \le r \le 1} \, \frac{f(x)}{f(r x + 1 - r)} = \frac{f(0)}{f(1-\gamma)} = \frac{1}{f(1-\gamma)}
\ee
\end{lemma}

\medskip
Lemma \ref{lem:f-bound} will be proved in Section \ref{pf:lem:f-bound}.
Recall (\ref{def:eps1}) and define
\be\label{def:z}
z_i = q_i - \epsilon_i = r p_i + (1-r) \frac{1}{d} 
\ee
Then Lemma \ref{lem:f-bound} implies that for each $i=1,\dots,d$,
\be\label{ineq:f1}
p_i d \, \log (p_i d) - p_i d + 1  = f(p_i d) \le \frac{1}{f(1-\gamma)} \, f(z_i d)
\ee
Therefore from (\ref{ent.cond2}) it follows that
\be\label{ent.cond3}
\sum_{i=1}^d \bigg(z_i d \, \log (z_i d) - z_i d + 1 \bigg) =
\sum_{i=1}^d f(z_i d) \ge h f(1-\gamma)
\ee

We will use the Fannes inequality \cite{Fannes73}, \cite{Audenaert07} to bound the difference between
the entropies of $z_i$ and $q_i$:
\be\label{bound.Fannes}
\Big| -\sum z_i \log z_i + \sum_i q_i \log q_i \Big| \le \epsilon_{m} \, ( \log d + \log \frac{1}{\epsilon_m} )
\ee
where
\be
\epsilon_{m} = \sum_{i=1}^d  |z_i - q_i|
= \sum_{i=1}^d | \epsilon_i | \le  t \, d \, \sqrt{\frac{d \log n}{n}}
\ee
Define
\be
\eta = d \, \epsilon_{m} \, ( \log d + \log \frac{1}{\epsilon_m} )
\ee
Note that for all $d$ and $t$, $\epsilon_{m} \rightarrow 0$ as $n \rightarrow \infty$, and
hence also $\eta \rightarrow 0$ as $n \rightarrow \infty$.
\medskip
From (\ref{bound.Fannes}) and (\ref{ent.cond3}) we deduce
\be\label{ent.cond4}
\sum_{i=1}^d f(q_i d) \ge h f(1-\gamma)
- \eta
\ee

\medskip
To summarize what we have shown so far: if $\rho \in J$ has eigenvalues
$(q_1,\dots,q_d)$ then (\ref{ent.cond4}) holds.
Thus we may upper bound the probability (\ref{uppr1}) by the probability
of the state $\rho$ satisfying the inequality (\ref{ent.cond4}). Since this event
depends only on the eigenvalues of $\rho$, we obtain
\be\label{upper3}
G^*(\sigma_{d n})(J) \le \mu_{d,n} \bigg\{ q \,:\, \sum_{i=1}^d f(q_i d) > h f(1-\gamma)
- \eta
\bigg\}
\ee
This probability is estimated using the bound (\ref{mu3a}): given a positive number $x \le d \log d$,
define 
\be\label{def:Mh}
M_d(x) = \inf_{q \in {\Delta}_d} \Big\{ \sum_{i=1}^d F(q_i d) \,:\, \sum_{i=1}^d f(q_i d) \ge x \Big\}
\ee
where $F(x) = - \log x + x -1$ as defined in (\ref{def:F}).
Then from (\ref{upper3}) and (\ref{mu3a}) we deduce
\be\label{upper4}
({\cal P}_{d,n} \times \sigma_n)(E) = G^*(\sigma_{d n})(J)
 \le  \exp \Big[d^2 \log n - \log (d-1)! - (n-d) M_d\Big(h f(1-\gamma) - \eta\Big) \Big]
\ee
The next Lemma gives a lower bound for $M_d(x)$ which is not optimal
but is sufficient for our purposes.

\begin{lemma}\label{lem:M(h)}
The function $M_d(x)$ is increasing.
Suppose that $2 e^2 \le x \le d \log d$. Then
\be\label{M-asymp}
M_d(x) \ge \log (x-1) - \log (2 e^2 - 1)
\ee
\end{lemma}

Lemma \ref{lem:M(h)} will be proved in Section \ref{pf:lem:M(h)}.
Applying (\ref{M-asymp}) to (\ref{upper4}) gives
\be\label{upper5}
({\cal P}_{d,n} \times \sigma_n)(E) 
 \le  \exp \bigg[d^2 \log n - \log (d-1)! - (n-d) \log \Big(\frac{h f(1-\gamma) - \eta - 1}{2 e^2 - 1}\Big) \bigg]
\ee
where $h$ is assumed to satisfy the bounds
\be\label{upper6}
2 e^2 \le h f(1-\gamma) - \eta \le d \log d
\ee

\subsection{The lower bound for $Prob(E)$ }
First we write
\bee
({\cal P}_{d,n} \times \sigma_n)(E) &=& \mathbb{E}_{\Phi}[\sigma_n(z \,:\, \Phi^C(z z^*) \in J)] \\
& \ge & \mathbb{E}_{\Phi}[1_{C_{d,n} \cap T} \, \sigma_n(z \,:\, \Phi^C(z z^*) \in J)]
\eee
where $\mathbb{E}_{\Phi}$ denotes expectation over ${\cal R}_d(n)$ with respect to the 
measure ${\cal P}_{d,n}$, and $1_{C_{d,n} \cap T}$ is the characteristic function of the event
${C_{d,n} \cap T}$.
Given that $\Phi \in C_{d,n}$ there is a state $v \in \mathbb{C}^n$ such that
\be
S(\Phi^C(v v^*)) \le \log d - \frac{h}{d}
\ee
Since $\tu(\Phi^C(v v^*)) \subset J$ it follows that
\be\label{ineq3a}
({\cal P}_{d,n} \times \sigma_n)(E) \ge 
\mathbb{E}_{\Phi}[1_{C_{d,n} \cap T} \, \sigma_n(z \,:\, \Phi^C(z z^*) \in \tu(\Phi^C(v v^*)))]
\ee
Applying Lemma \ref{lem:uniform-lwr-bnd} to (\ref{ineq3a}) gives
\be\label{ineq3b}
({\cal P}_{d,n} \times \sigma_n)(E) & \ge & \beta \, \bigg( 1 - \gamma \bigg)^{n-1} \,
\mathbb{E}_{\Phi}[1_{C_{d,n} \cap T}] \\ 
& = & \beta \,\bigg( 1 - \gamma \bigg)^{n-1} \, {\cal P}_{d,n}(C_{d,n} \cap T) \\
& \ge & \beta \,\bigg( 1 - \gamma \bigg)^{n-1} \, ({\cal P}_{d,n}(C_{d,n}) - {\cal P}_{d,n}(T^c)) 
\ee

\subsection{Combining the bounds for $Prob(E)$ and finishing the proof}
Putting together the upper and lower bounds for $({\cal P}_{d,n} \times \sigma_n)(E)$
and using Lemma \ref{lem:prob-Tc} produces the following bound: for all $d \ge 3$,
for all $b > \sqrt{3}$ and $t \ge b+4$, for all $0 < \gamma < 1$, for $h,d$ satisfying (\ref{upper6}),
and for $n$ sufficiently large
\be\label{ineq:16}
{\cal P}_{d,n}(C_{d,n}) & \le & {\cal P}_{d,n}(T^c) + \frac{1}{\beta} \, \bigg( \frac{1}{1 - \gamma} \bigg)^{n-1} \, 
({\cal P}_{d,n} \times \sigma_n)(E) \nonumber \\
& \le & \frac{2 \, d}{(d-1)!} \,\, \exp[-\alpha d^2 \, \log n] + \frac{1}{\beta} \, \bigg( \frac{1}{1 - \gamma} \bigg)^{n-1} \, 
\exp [d^2 \log n - \log (d-1)! - (n-d) \log \tilde{h} ] \nonumber \\
&= & \frac{2 \, d}{(d-1)!} \,\, \exp[-\alpha d^2 \, \log n] \nonumber \\
&& \hskip 0.3in+ \frac{1-\gamma}{\beta (d-1)!} \,
\exp[d^2 \log n + d \log \tilde{h}  - n  \log (1 - \gamma) \tilde{h}]
\ee
where $\tilde{h} = (h f(1-\gamma) - \eta -1)/(2 e^2 - 1)$. 
Define 
\be\label{def:h-min}
h_{\min} = \frac{2 e^2 - \gamma}{(1-\gamma) f(1-\gamma)}
\ee
(note that $h_{\min}$ satisfies the lower bound in (\ref{upper6})).
As $n \rightarrow \infty$ the parameter $\eta$ approaches zero,
and therefore for $h > h_{\min}$ the second term on the right side of (\ref{ineq:16}) is controlled by the factor
\be\label{ineq:17}
\exp\bigg[ - n \log \frac{(1 - \gamma)( h f(1-\gamma) -1)}{2 e^2 - 1}\bigg] =
\bigg( \frac{h f(1-\gamma) -1}{h_{\min} f(1-\gamma) -1} \bigg)^{-n}
\ee
The first factor on the right side of (\ref{ineq:16}) approaches zero as $n \rightarrow \infty$,
therefore (\ref{ineq:17}) implies that for $h > h_{\min}$,
\be
{\cal P}_{d,n}(C_{d,n}) \rightarrow 0 \quad {\rm as} \,\, n \rightarrow \infty
\ee

\medskip
\noindent \fbox{Summary and conclusion} We have shown that for any $0 < \gamma < 1$,
for $h > h_{\min}$ as defined in (\ref{def:h-min}), for any $b > \sqrt{3}$ and $t \ge b+4$,
for any $d \ge 3$ satisfying $d \log d > h f(1-\gamma)$ (this comes from the second inequality
in (\ref{upper6})), there is $N < \infty$ such that for all $n \ge N$ we have
${\cal P}_{d,n}(C_{d,n}) < 1$. In this case we also have ${\cal P}_{d,n}(C_{d,n}^c) > 0$,
and thus a guarantee that the set $C_{d,n}^c$ is non-empty.
Referring to (\ref{def:C-d,n}), this means that there exists a random unitary channel
$\Phi$ such that
\be
S_{\min}(\Phi) > \log d - \frac{h}{d}
\ee

\subsection{Optimizing the bounds for $Prob(E)$ 
and the proof of Proposition \ref{prop:min-dims}}\label{pf:prop:min-dims}
First consider the value $h_{\min}$ defined in (\ref{def:h-min}).
Varying $\gamma$ shows that the right side achieves its minimum value
at $\gamma = 0.72$. In order to achieve a counterexample we need 
$\log d \ge 2 h$, so this implies the existence of counterexamples for all
$d \ge d_0$ with
\be\label{optim:1}
d_0 = \exp [2 h_{\min} + 1] \simeq \exp[276]
\ee

\medskip
In order to get  a better estimate of $d_{\min}$, we return to the bound
(\ref{upper4}) and look for the smallest
value of $d$ satisfying
\be\label{prob-ineq1}
M_d\Big(\frac{f(1-\gamma)}{2} \log d \Big) + \log (1 - \gamma) > 0
\ee
For $n$ sufficiently large this will yield a counterexample.
This is a straightforward numerical problem:
for each $\gamma$ we find the smallest $d$ so that
\be
 - \log (1 - \gamma) < && \hskip-0.15in \inf_{z > 1} \{ - \log z - (d-1) \log \frac{d-z}{d-1} : \nonumber \\
&& \hskip0.1in z \log z + (d-z) \log \frac{d-z}{d-1} = \frac{f(1-\gamma)}{2} \log d \}
\ee
and then minimize over $\gamma$. The solution occurs at $\gamma = 0.762$ and yields
$d_0 = 38578$. This also proves the first statement in Proposition \ref{prop:min-dims}.

\medskip
For the second statement we estimate the smallest value of $n$ which yields
${\cal P}_{d,n}(C_{d,n}) < 1$. Using the values $b=2$, $t=6$, $\gamma=0.762$, and with
$d = 50,000$, crude numerical estimates show that we can achieve this with $n = d^7$. 
This proves the second statement in Proposition \ref{prop:min-dims}.

\medskip
For the third statement, we note from Lemma \ref{lemma01} that for any random unitary channel $\Phi$
\be
\Delta S_{\max} \ge 2 S_{\min}(\Phi) - 2 \log d + \frac{\log d}{d}
\ee
Thus for every $\Phi \in C_{d,n}^c$ we have
\be\label{optim:2}
\Delta S_{\max} \ge \frac{\log d - 2 h}{d}
\ee
For a fixed value $h$,
the right side of (\ref{optim:2}) achieves its maximum value
when $d = [\exp (2 h + 1)]$, and this maximum value is $1/d$.
Numerical calculation shows that we can achieve
$M_d(f(1-\gamma) h) + \log (1 - \gamma) > 0$ using the values $\gamma=0.762$,
$h = \log (38590)/2$ and $d = [\exp (2 h + 1)]$, and then $1/d$
yields the lower bound for $\Delta S_{\max}$ stated in Proposition \ref{prop:min-dims}.

\section{Proofs of Lemmas}
\subsection{Proof of Lemma \ref{lemma01}}\label{pf:lemma01}
First, note that for any unit vectors $\{|\psi_k\rangle\}$ and probability distribution
$\{p_k\}$, 
\begin{align}
S\left(\sum_k p_k|\psi_k \rangle\langle\psi_k|\right)
\leq 
- \sum_k p_k \ln p_k.
\end{align}
Let $|\hat{\psi}\rangle$ be the maximally entangled state. Then
\begin{align}
(\Phi \otimes \overline{\Phi}) (|\hat{\psi}\rangle\langle\hat{\psi}| )
&= \sum_{i,j=1}^d w_iw_j \; U_i \otimes \overline{U_j} 
|\hat{\psi}\rangle\langle\hat{\psi}| 
U_i^\ast \otimes  U_j^T  \\
&= \left(\sum_{i=1}^d w_i^2 \right) |\hat{\psi}\rangle\langle\hat{\psi}|
+ \sum_{i \not= j} w_iw_j \; U_i \otimes \overline{U_j} 
|\hat{\psi}\rangle\langle\hat{\psi}| 
U_i^\ast \otimes  U_j^T 
\end{align}
where we used the identity $U_i \otimes \overline{U_i} 
|\hat{\psi}\rangle\langle\hat{\psi}| 
U_i^\ast \otimes  U_i^T  = |\hat{\psi}\rangle\langle\hat{\psi}|$
for all $i$.
Hence,
\begin{align}
S\left( (\Phi \otimes \overline{\Phi}) (|\hat{\psi}\rangle\langle\hat{\psi}| )\right)
\leq - \left(\sum_{i=1}^d w_i^2 \right) \ln \left(\sum_{i=1}^m w_i^2 \right)
- \sum_{i \not= j} w_iw_j \ln ( w_iw_j).
\label{tensorbound2}
\end{align}

Write $p=\sum_{i=1}^d w_i^2$ and
then $\sum_{i \not= j}w_iw_j = 1-p$. Hence
\begin{align}
S\left( (\Phi \otimes \overline{\Phi}) (|\hat{\psi}\rangle\langle\hat{\psi}| )\right)
\leq - p \ln p + \sup \bigg\{
- \sum_{k=1}^{d^2-d} v_k \ln v_k \,:\, v_k \ge 0, \,\, \sum_{k=1}^{d^2-d} v_k = 1-p \bigg\}
\label{tensorbound3}
\end{align}
The supremum on the right side of (\ref{tensorbound3}) is achieved with $v_k = (1-p)/(d^2-d)$
for all $k$, hence
\begin{align}
S\left( (\Phi \otimes \overline{\Phi}) (|\hat{\psi}\rangle\langle\hat{\psi}| )\right)
\leq
h(p) = -p\ln p - (1-p) \ln \left(\frac{1-p}{d^2 -d}\right),
\end{align}
where $1/d \leq p \leq1$. However,
\begin{align}
h^\prime(p) = -\ln p +\ln(1-p) - \ln (d^2 -d) 
= -\ln (pd) + \ln\left(\frac{1-p}{d-1}\right)<0
\end{align}
for all $d$.
This implies that 
the above upper bound $h(p)$ is maximized when $p=1/d$ and the maximum is
\begin{align}
\frac{1}{d}\ln d
+ \left(1-\frac{1}{d} \right) \ln (d^2)= 2 \ln d - \frac{1}{d} \ln d.
\end{align}

\subsection{Proof of Lemma \ref{lem:bound-mu}}\label{pf:lem:bound-mu}
By dropping the terms $(w_i-w_j)^2$ in (\ref{mu-density}) we get
\be\label{lem:mu1}
\mu_{d,n}(A) \le Z(n,d)^{-1} \, \int_{A} 
\prod_{i=1}^d w_i^{n-d} \, \delta\bigg(\sum_{i=1}^d w_i - 1\bigg) \, [dw]
\ee
Applying (\ref{Z-est1}) to (\ref{lem:mu1}) leads to
\be\label{lem:mu3}
\mu_{d,n}(A) \le  \int_{A} \exp[ d^2 \log n + (n-d) \, \sum_{i=1}^d \log (d w_i)] 
\, \delta\bigg(\sum_{i=1}^d w_i - 1\bigg) \, [dw]
\ee
Noting that 
\be
\sum_{i=1}^d F(d w_i) = - \sum_{i=1}^d \log (d w_i)
\ee
the result follows.

\subsection{Proof of Lemma \ref{lem1}}\label{pf:lem1}
For any event $A \subset {\cal M}_d$,
\be\label{lem1pf:1}
H^*({\cal P}_{d,n} \times \sigma_n)(A) & = & ({\cal P}_{d,n} \times \sigma_n)(H^{-1}(A)) \nonumber \\
& = & \int d \sigma_{n}(z) \, {\cal P}_{d,n} \Big\{ \Phi \,:\, \Phi^C(z z^*) \in A \Big\}
\ee
A random unitary channel $\Phi$ is determined by the coefficients $\{w_1,\dots,w_d\}$ and the unitary matrices
$\{U_1,\dots,U_d\}$.
Given a unitary matrix $V$ define the transformation 
$T_V : {\cal R}_d(n) \rightarrow {\cal R}_d(n)$ by
\be\label{lem1pf:4}
T_V \,:\, \{w_1,\dots,w_d; U_1,\dots,U_d \} \rightarrow  \{w_1,\dots,w_d; U_1 V ,\dots,U_d V \}
\ee
The measure ${\cal P}_{d,n} =  \nu_{d,n} \times H_n \times \cdots \times H_n$ contains the product
of $d$ independent copies of Haar measure $H_n$ on the group ${\cal U}_n$. Since Haar measure is invariant under
group multiplication, for any event $C \subset {\cal R}_d(n)$ we have
\be\label{lem1pf:5}
{\cal P}_{d,n}(C) & = &
\int_C d \nu_{d,n}(w_1,\dots,w_d) \, d H_n(U_1) \cdots d H_n(U_n) \nonumber \\
& = & \int_{T_V(C)} d \nu_{d,n}(w_1,\dots,w_d) \, d H_n(U_1) \cdots d H_n(U_n) \nonumber \\
& = & {\cal P}_{d,n}(T_V(C))
\ee
Thus in particular for any $z \in {\cal V}_{n}$,
\be\label{lem1pf:6}
{\cal P}_{d,n} \Big\{ \Phi \,:\, \Phi^C(z z^*) \in A \Big\} = {\cal P}_{d,n} \Big\{ \Phi \,:\, \Phi^C(V z (V z)^*) \in A \Big\}
\ee
Since ${\cal U}_n$ acts transitively on ${\cal V}_{n}$, (\ref{lem1pf:6}) shows that the probability is independent of $z$.
Hence from (\ref{lem1pf:1}) we obtain that for any fixed $z_0 \in {\cal V}_{n}$,
\be\label{lem1pf:7}
H^*({\cal P}_{d,n} \times \sigma_n)(A) = {\cal P}_{d,n} \Big\{ \Phi \,:\, \Phi^C(z_0 z_0^*) \in A \Big\}
\ee

\medskip
For a given channel $\Phi$ the $d \times d$ matrix $\Phi^C(z_0 z_0^*)$ can be written in terms of 
a $n \times d$ matrix $K(\Phi)$ as follows:
\be\label{lem1pf:2}
\Phi^C(z_0 z_0^*) = K(\Phi)^* K(\Phi), \quad
 K(\Phi) = \bmx \sqrt{w_1} v_1 & \cdots & \sqrt{w_d} v_d \emx
\ee
where for $i=1,\dots,d$
\be\label{lem1pf:3}
v_i = \overline{U_i z_0}
\ee
Thus (\ref{lem1pf:7}) can be written as
\be\label{lem1pf:7a}
H^*({\cal P}_{d,n} \times \sigma_n)(A) = {\cal P}_{d,n} \Big\{ \Phi \,:\, K(\Phi)^* K(\Phi) \in A \Big\}
\ee

\medskip
Recall from (\ref{def:G}) that $G(z) = M(z)^* M(z)$ where $z \in {\cal V}_{nd}$ and where
the $n \times d$ matrix $M(z)$ has entries
\be\label{lem1pf:8}
M_{ij}(z) = z_{(i-1)d + j}, \quad i=1,\dots n, \,\, j=1,\dots,d
\ee
It follows that for any event $A \subset {\cal M}_d$,
\be\label{lem1pf:9}
G^*(\sigma_{dn})(A) = \sigma_{dn} \Big\{z \,:\, M(z)^* M(z) \in A \Big\}
\ee

\medskip
We wish to prove that $H^*({\cal P}_{d,n} \times \sigma_n)(A) = G^*(\sigma_{dn})(A)$ for every event $A \subset {\cal M}_d$.
Comparing (\ref{lem1pf:9}) and (\ref{lem1pf:7a}), it is sufficient to show that
the $d \times d$ matrices $K(\Phi)$ and $M(z)$ have the same distribution.
We will do this by showing that the columns of $K(\Phi)$ and $M(z)$
have the same joint distribution.
Before showing the result, we need the following observation on a normally distributed vector.

\medskip
Let $Z_1,\dots,Z_m$ be IID complex valued normal random variables with mean zero and variance one.
Let $R = \sqrt{\sum_{i=1}^m |Z_i|^2}$ and define the vector
\be\label{lem1pf:12}
\xi = \frac{1}{R} \bmx Z_1 \cr \vdots \cr Z_m \emx
\ee
Then $R,\xi$ are independent, $\xi$ is a random pure state in ${\cal V}_m$, and $R$ has density
\be\label{lem1pf:13}
P(r) \propto e^{-r^2/2} \, r^{2m-1}
\ee
This result may be easily seen by transforming the joint density for $Z_1,\dots,Z_m$ to polar coordinates:
\be\label{lem1pf:14}
(2 \pi)^{-m} \, \prod_{i=1}^m e^{-|z_i|^2/2} \, d^2 z_i = \pi^{-m} e^{-r^2/2} \, r^{2m-1} \, d r \, d \Omega
\ee
where $d \Omega$ is the uniform measure on $S^{2m-1}$.

\medskip
We look at $M(z)$ first.
Let $\{Z_{ij}\}$ ($i=1,\dots,n$, $j=1,\dots,d$) be a collection of
IID complex valued normal random variables with mean zero and variance one, arranged into a $n \times d$ matrix
$Z$ as in (\ref{lem1pf:8}). Applying the previous observation to $Z$ and also to each column of $Z$ yields
\be\label{lem1pf:15}
Z = R \, M = \bmx R_1 \, \xi_1 & \cdots & R_d \, \xi_d \emx
\ee
Here $\{\xi_1,\dots,\xi_d\}$ are IID random unit vectors in ${\cal V}_n$, and $M$ is a random unit vector
in ${\cal V}_{nd}$. The vectors $\{\xi_1,\dots,\xi_d\}$ are independent of the 
numbers $R_1,\dots,R_d$. Also $R^2 = R_1^2 + \cdots + R_d^2$, hence $\{\xi_1,\dots,\xi_d\}$ are also
independent of $R$. Dividing by $R$, 
$M(z)$, a random unit vector in ${\cal V}_{nd}$, can be reconstructed as 
\be\label{lem1pf:16}
M(z) = \bmx \sqrt{Y_1} \, \xi_1 & \cdots & \sqrt{Y_d} \, \xi_d \emx, \quad Y_i = \frac{R_i^2}{R^2}
\ee
Note that $Y_1,\dots,Y_d$
have the joint distribution $\nu_{d,n}$, and are independent of $\{\xi_1,\dots,\xi_d\}$.

\medskip
Next, turning into $K(\Phi)$,
recall that 
\be
K(\Phi) = \bmx \sqrt{w_1} v_1 & \cdots & \sqrt{w_d} v_d \emx
\ee
where $v_i = \overline{U_i z_0}$. Since the
unitaries $U_i$ are independently and uniformly selected (this is part of the definition of the measure
${\cal P}_{d,n}$), it follows that the vectors $\{v_i\}$
are IID random unit vectors in ${\cal V}_n$. Furthermore the coefficients $\{w_i\}$ have the joint distribution
$\nu_{d,n}$. 
This verifies our claim.

\subsection{Proof of Lemma \ref{lem:prob-Tc}}\label{pf:lem:prob-Tc}
Define the following subset of ${\cal R}_d(n) \times {\cal V}_n$:
\be\label{def:K}
K = \{ (\Phi,z) \,:\, \Phi^C(z z^*) \notin B_d(n) \} = H^{-1}(B_d(n)^c)
\ee
where the map $H$ was defined in (\ref{def:H}). Then
\be
({\cal P}_{d,n} \times \sigma_{n})(K) & = & \mathbb{E}_{\Phi} [\sigma_n(z \,:\,
\Phi^C(z z^*) \notin B_d(n)) ] \nonumber \\
& \ge & \mathbb{E}_{\Phi} [1_{T^c} \,\, \sigma_n(z \,:\,
\Phi^C(z z^*) \notin B_d(n)) ] 
\ee
where $\mathbb{E}_{\Phi}$ denotes expectation over ${\cal R}_d(n)$ with respect to the 
measure ${\cal P}_{d,n}$, and $1_{T^c}$ is the characteristic function of the event
$T^c$. Note that if $\Phi \in T^c$ then 
$\sigma_n(z \,:\,
\Phi^C(z z^*) \notin B_d(n)) \ge1/2$, hence
\be
({\cal P}_{d,n} \times \sigma_{n})(K)
& \ge & \frac{1}{2} \, \mathbb{E}_{\Phi} [1_{T^c}] = 
\frac{1}{2} \, {\cal P}_{d,n}(T^c)
\ee

\medskip
Furthermore from Lemma \ref{lem1} it follows that
\be
({\cal P}_{d,n} \times \sigma_{n})(K) & = & ({\cal P}_{d,n} \times \sigma_{n})(H^{-1}(B_d(n)^c)) \nonumber \\
& = & H^*({\cal P}_{d,n} \times \sigma_{n})(B_d(n)^c) \nonumber \\
& = & G^*(\sigma_{dn})(B_d(n)^c)
\ee
Combining these bounds shows that
\be\label{prob-Tc-2}
{\cal P}_{d,n}(T^c) & \le & 2 \, G^*(\sigma_{d n})({\cal V}_d(n)^c) \nonumber \\
&=& 2 \, \mu_{d,n} \{(q_1,\dots,q_d) \,:\, |q_i - 1/d | > b \sqrt{\frac{\log n}{n}} \,\, {\rm some} \,\, i =1, \dots,d \} \nonumber \\
& = & 2 \mu_{d,n}\Big(\bigcup_{i=1}^d L_i\Big)
\ee
where the events $L_i$ are defined by
$L_i = \{(q_1,\dots,q_d) \,:\, |q_i - 1/d | > b \sqrt{\log n/n} \}$. Thus we have
\be\label{prob-Tc-2a}
{\cal P}_{d,n}(T^c)  \le  
2 \sum_{i=1}^d  \mu_{d,n}(L_i) = 2 \, d \, \mu_{d,n}(L_i)
\ee

\medskip
We use the bound (\ref{mu3}) of Corollary  \ref{cor2} with $t = b \sqrt{\log n/n}$
to estimate $\mu_{d,n}(L_i)$. In addition we assume that $n$ is large enough so that
\be
 d \, t = d \, b \, \sqrt{\frac{\log n}{n}} \le 1
\ee
and hence
\be
\frac{n-d}{2} d^2 t^2 - \frac{n-d}{3!} d^3 t^3 \ge \frac{n-d}{3} d^2 t^2
\ee
Thus (\ref{prob-Tc-2}) gives
\be\label{prob-Tc-3}
{\cal P}_{d,n}(T^c) & \le &  2 \, d \, 
\exp \Big[d^2 \log n - \log (d-1)! - \frac{n-d}{3} d^2 b^2 \frac{\log n}{n} \Big] \nonumber \\
& = & \frac{2 \, d}{(d-1)!} \, 
\exp \Big[- d^2 \log n \bigg( \frac{b^2 (n-d)}{3n} - 1\bigg) \Big]
\ee

\subsection{Proof of Lemma \ref{lem:uniform-lwr-bnd}}\label{pf:lem:uniform-lwr-bnd}
This result relies on several properties of random states. We will switch to Dirac
bra and ket notation throughout this Section, as it lends itself well to the arguments used in the proof.
To set up the notation,
let $| \psi \ket$ be a fixed state in ${\cal V}_n$, and let $| \theta \ket$ be a random pure state
in ${\cal V}_n$, with probability distribution $\sigma_n$. Without loss of generality
we assume that a basis is chosen so that  $| \psi \ket = (1,0,\dots,0)^T$.
We write
$x = \bra \psi | \theta \ket$, and let $| \phi \ket$ be the state orthogonal to
$| \psi \ket$ such that
\be\label{def:phi}
| \theta \ket = x \, | \psi \ket + \sqrt{1 - |x|^2} \, | \phi \ket
\ee
Thus $| \phi \ket$ is also a random state, defined by its relation to the uniformly random state
$| \theta \ket$ in (\ref{def:phi}). The following results are proved in Appendix B.

\begin{prop}\label{random-states}
$x$ and $| \phi \ket$ are independent. $| \phi \ket$ is a random vector
in ${\cal V}_{n-1}$ with distribution $\sigma_{n-1}$.
For all $0 \le t \le 1$
\be\label{Prop:eqn1}
\sigma_n\{ | \theta \ket \,:\, |\bra \psi | \theta \ket| = |x| > t \} = (1 - t^2)^{n-1}
\ee
\end{prop}

Proposition \ref{random-states} implies that
as $n \rightarrow \infty$ the overlap $x = \bra \psi | \theta \ket$ becomes concentrated around zero.
In other words, with high probability a randomly chosen state will be almost
orthogonal to any fixed state. As a  consequence, from (\ref{def:phi}) it follows that
$| \phi \ket$ will be almost equal to $| \theta \ket$. This statement is made precise
by noting that
\be
\| | \theta \ket - | \phi \ket \|_{\infty} = |\bra \psi | \theta \ket|
\ee
Then (\ref{Prop:eqn1}) immediately implies that
\be\label{prop1a}
\sigma_n(| \theta \ket \,:\, \| | \theta \ket - | \phi \ket \|_{\infty} > t) = (1 - t^2)^{n-1}
\ee

\medskip
The second property relies on the particular form of the random unitary channel, or more
precisely on the form of the complementary channel $\Phi^C$. Roughly, this property says that
for any fixed random unitary channel $\Phi$ and random state $| \theta \ket$, with high probability
the norm of the matrix $\Phi^C(| \theta \kb \psi |)$ is small,
and approaches zero as $n \rightarrow \infty$. We will prove the following bound:
for any $\Phi \in {\cal R}_d(n)$, and for all $0 \le t \le 1$,
\be\label{prop2}
\sigma_n(| \theta \ket \,:\, \| \Phi^C(| \theta \kb \psi |) \|_2 > t) \le d^2 \, (1 - t^2)^{n-1}
\ee

\medskip
As a first step toward deriving (\ref{prop2}), note that for any states
$| u \ket$ and $| v \ket$,
\be\label{prop2a}
\| \Phi^C(| u \kb v |) \|_2 = 
\left(\sum_{k,l =1}^d w_k w_l 
 |\bra v | U_l^\ast U_k |u \ket |^2
 \right)^{\frac{1}{2}}
\leq \max_{k,l} |\bra v | U_l^\ast U_k |u\ket|,
\ee
In particular this implies that 
\be\label{prop1b}
\| \Phi^C(| u \kb v |) \|_2 \le \max \{\| | u \ket \|_{\infty}, \, \| | v \ket \|_{\infty} \}
\ee
To derive (\ref{prop2}) we apply (\ref{prop2a}) with $u = \theta$ and $v = \psi$ and deduce that
\be\label{prop2b}
\sigma_n( | \theta \ket \,:\, \| \Phi^C(| \theta \kb \psi |) \|_2 > t) & \le &
\sigma_n( | \theta \ket \,:\, \max_{k,l} |\bra \psi | U_l^\ast U_k | \theta \ket| > t) \nonumber \\
& \le & d^2 \,
\sigma_n( | \theta \ket \,:\, |\bra \psi | U_l^\ast U_k | \theta \ket| > t) \nonumber \\
& = & d^2 \, (1 - t^2)^{n-1}
\ee
where the last equality follows from (\ref{Prop:eqn1}).

\medskip
With these ingredients in place the proof of Lemma \ref{lem:uniform-lwr-bnd} can proceed.
By assumption $\Phi$ is a random unitary channel belonging to the typical set $T$, and
$\rho = \Phi^C(|\psi \kb \psi|)$ is some state in ${\rm Im}(\Phi^C)$.
Let $| \theta \ket$ be a random input state, then as in (\ref{def:phi}) we write
\be\label{def:phi2}
| \theta \ket = x \, | \psi \ket + \sqrt{1 - |x|^2} \, | \phi \ket \nonumber
\ee
It follows that
\be\label{tube2}
| \theta \kb \theta | = |x|^2 \, | \psi \kb \psi | + (1 - |x|^2) \, | \phi \kb \phi |
+ \sqrt{1 - |x|^2} \, (x \, | \psi \kb \phi | + \overline{x} \, | \phi \kb \psi |)
\ee
Write $r = |x|^2$, then (\ref{tube2}) yields
\be\label{tube3}
\Phi^C(|\theta\rangle\langle\theta|) 
&& \hskip-0.2in - \left(
r \Phi^C(|\psi\rangle\langle\psi|) + 
(1-r)\frac{1}{d}I\right) \nonumber \\
&&\quad = (1-r) \left( \Phi^C |\phi\rangle\langle\phi|)- \frac{1}{d}I\right)
+\sqrt{r(1-r)}
\Phi^C\left( e^{i \xi} \, |\psi\rangle\langle\phi| + e^{- i \xi} \, |\phi\rangle\langle\psi| \right)
\ee
where $\xi$ is the phase of $x$. Since $r \le 1$ this implies
\be\label{tube4}
\bigg\| \Phi^C(|\theta\rangle\langle\theta|) 
 - \left(
r \Phi^C(|\psi\rangle\langle\psi|) + 
(1-r)\frac{1}{d}I\right) \bigg\|_{\infty} 
 \le \bigg\|  \Phi^C (|\phi\rangle\langle\phi|)- \frac{1}{d}I \bigg\|_{\infty} +
\bigg\| \Phi^C (|\psi\rangle\langle\phi|) \bigg\|_{\infty}
\ee
Referring to the definition (\ref{def:tube}) of $\tu (\rho)$,
recall that $\Phi^C(|\theta\rangle\langle\theta|)$ belongs to  $\tu(\rho)$ if and only if for some $r$
satisfying $\gamma \le r \le 1$,
\be\label{tube5}
\bigg\| \Phi^C(|\theta\rangle\langle\theta|) 
- \left(
r \Phi^C(|\psi\rangle\langle\psi|) + 
(1-r)\frac{1}{d}I\right) \bigg\|_{\infty} \le t \, \sqrt{\frac{d \log n}{n}}
\ee
(the set $Y(\rho)$ defined in (\ref{def:Y}) is closed so the infimum in (\ref{def:tube}) is achieved).
Define the following three events in ${\cal V}_n$:
\be\label{def:events}
A_1 & = & \{| \theta \ket \,:\, r = |\bra \psi | \theta \ket|^2 \ge \gamma \} \\
A_2 &= & \bigg\{| \theta \ket \,:\, \bigg\|  \Phi^C (|\phi\rangle\langle\phi|)- \frac{1}{d}I \bigg\|_{\infty}
\le 2 \, \sqrt{\frac{d \log d}{n}} +  b \, \sqrt{\frac{d \log n}{n}} \bigg\} \\
A_3 &= & \bigg\{| \theta \ket \,:\, \bigg\| \Phi^C (|\psi\rangle\langle\phi|) \bigg\|_{\infty} \le 2\,
\sqrt{\frac{d \log d}{n}} \bigg\}
\ee
Since $t \ge b+4$ and $d \le n$, it follows from (\ref{tube4}) and (\ref{tube5}) that
\be
A_1 \cap A_2 \cap A_3 \subset \{| \theta \ket \,:\, 
\Phi^C(|\theta\rangle\langle\theta|) \in \tu(\rho) \}
\ee
Furthermore by Proposition \ref{random-states}, $A_1$ is independent of $A_2$ and $A_3$, hence
\be\label{tube6}
\sigma_n(\Phi^C(|\theta\rangle\langle\theta|) \in \tu(\rho)) \ge
\sigma_n(A_1 \cap A_2 \cap A_3) = 
\sigma_n(A_1) \, \sigma_n(A_2 \cap A_3)
\ee

\medskip
Proposition \ref{random-states} immediately yields
\be
\sigma_n(A_1) = (1 - \gamma)^{n-1}
\ee
From (\ref{tube6}) this gives
\be\label{tube7}
\sigma_n(\Phi^C(|\theta\rangle\langle\theta|) \in \tu(\rho)) \ge
(1 - \gamma)^{n-1} \,(1 - \sigma_n(A_2^c) - \sigma_n(A_3^c))
\ee
In order to bound $\sigma_n(A_3^c)$ we first use (\ref{prop1b}) to deduce
\be\label{tube8}
\| \Phi^C (|\psi\rangle\langle\phi|) \|_{\infty} \le
\| \Phi^C (|\psi\rangle\langle\phi|) \|_{2} \le
\| \Phi^C (|\psi\rangle\langle\theta|) \|_{2} + \| | \theta \ket - | \phi \ket \|_{\infty}
\ee
Thus
\be\label{tube9}
\sigma_n(A_3^c) & = & \sigma_n\bigg\{| \theta \ket \,:\, \bigg\| \Phi^C (|\psi\rangle\langle\phi|) \bigg\|_{\infty} >
2 \,\sqrt{\frac{d \log d}{n}} \bigg\} \nonumber \\
& \le & \sigma_n\bigg\{| \theta \ket \,:\, 
\| \Phi^C (|\psi\rangle\langle\theta|) \|_{2} + \| | \theta \ket - | \phi \ket \|_{\infty} >
2 \,\sqrt{\frac{d \log d}{n}} \bigg\} \nonumber \\
& \le & \sigma_n\bigg\{| \theta \ket \,:\, 
\| \Phi^C (|\psi\rangle\langle\theta|) \|_{2} > \sqrt{\frac{d \log d}{n}} \bigg\} +
\sigma_n\bigg\{| \theta \ket \,:\, \| | \theta \ket - | \phi \ket \|_{\infty} >
\sqrt{\frac{d \log d}{n}} \bigg\} \nonumber \\
& \le & (d^2 + 1) \, \Big(1 - \frac{d \log d}{n} \Big)^{n-1} \\
\ee
where the last inequality follows from (\ref{prop2b}) and (\ref{prop1a}).

\medskip
Turning now to $\sigma_n(A_2^c)$, note first that
\be\label{tube10}
\bigg\|  \Phi^C (|\phi\rangle\langle\phi|)- \frac{1}{d}I \bigg\|_{\infty} & \le &
\bigg\|  \Phi^C (|\phi\rangle\langle\phi|)- \Phi^C (|\theta\rangle\langle\theta|) \bigg\|_{\infty} +
\bigg\|  \Phi^C (|\theta\rangle\langle\theta|)- \frac{1}{d}I \bigg\|_{\infty} \nonumber \\
& \le &
\bigg\|  \Phi^C (|\phi\rangle\langle\phi|)- \Phi^C (|\theta\rangle\langle\theta|) \bigg\|_{2} +
\bigg\|  \Phi^C (|\theta\rangle\langle\theta|)- \frac{1}{d}I \bigg\|_{\infty} \nonumber \\
& \le & 2 \, \| | \theta \ket  - | \phi \ket \|_{\infty} +
\bigg\|  \Phi^C (|\theta\rangle\langle\theta|)- \frac{1}{d}I \bigg\|_{\infty}
\ee
where we used (\ref{prop1b}) for the last inequality. As in (\ref{tube9}) this gives
\be\label{tube11}
\sigma_n(A_2^c) & = & \sigma_n\bigg\{| \theta \ket \,:\, 
\bigg\|  \Phi^C (|\phi\rangle\langle\phi|)- \frac{1}{d}I \bigg\|_{\infty} >
2 \,  \sqrt{\frac{d \log d}{n}} + b \, \sqrt{\frac{d \log n}{n}} \bigg\} \nonumber \\
& \le & \sigma_n\bigg\{| \theta \ket \,:\, 
2 \, \| | \theta \ket  - | \phi \ket \|_{\infty} > 2 \,  \sqrt{\frac{d \log d}{n}} \bigg\} \nonumber \\
&&\hskip0.5in + \sigma_n\bigg\{| \theta \ket \,:\, 
\bigg\|  \Phi^C (|\theta\rangle\langle\theta|)- \frac{1}{d}I \bigg\|_{\infty} >
b \, \sqrt{\frac{d \log n}{n}} \bigg\} \nonumber \\
& \le & \bigg(1 - \frac{d \log d}{n} \bigg)^{n-1} +
\sigma_n\bigg\{| \theta \ket \,:\, 
\bigg\|  \Phi^C (|\theta\rangle\langle\theta|)- \frac{1}{d}I \bigg\|_{\infty} >
b \, \sqrt{\frac{d \log n}{n}} \bigg\}
\ee
where we used (\ref{prop1a}) for the last inequality. By assumption $\Phi \in T$, and 
therefore there is a set of input states $L$ with $\sigma_n(L) \ge 1/2$ such that
\be\label{tube12}
|\theta \ket \in L \Rightarrow 
\bigg\|  \Phi^C (|\theta\rangle\langle\theta|)- \frac{1}{d}I \bigg\|_{\infty} \le
b \, \sqrt{\frac{d \log n}{n}}
\ee
Thus
\be\label{tube13}
\sigma_n\bigg\{| \theta \ket \,:\, 
\bigg\|  \Phi^C (|\theta\rangle\langle\theta|)- \frac{1}{d}I \bigg\|_{\infty} >
b \, \sqrt{\frac{d \log n}{n}} \bigg\} \le
\sigma_n(L^c) \le \frac{1}{2}
\ee

\medskip
Putting together the bounds (\ref{tube7}), (\ref{tube9}), (\ref{tube11}) and (\ref{tube13}) we get
\be\label{tube14}
\sigma_n(\Phi^C(|\theta\rangle\langle\theta|) \in \tu(\rho))
 &\ge & 
(1 - \gamma)^{n-1} \,\bigg(1 - \sigma_n(A_2^c) - \sigma_n(A_3^c)\bigg) \nonumber \\
& \ge &  (1 - \gamma)^{n-1} \,\bigg(1 - \bigg(1 - \frac{d \log d}{n} \bigg)^{n-1} - \frac{1}{2}
- (d^2 + 1) \, \Big(1 - \frac{d \log d}{n} \Big)^{n-1}\bigg) \nonumber \\
& =&  (1 - \gamma)^{n-1} \,\bigg(\frac{1}{2} - (d^2 + 2) \, \Big(1 - \frac{d \log d}{n} \Big)^{n-1}\bigg)
\ee
This completes the proof, with 
\be
\beta = \bigg(\frac{1}{2} - (d^2 + 2) \, \Big(1 - \frac{d \log d}{n} \Big)^{n-1}\bigg)
\ee 

\subsection{Proof of Lemma \ref{lem:f-bound}}\label{pf:lem:f-bound}
It is clear that $f(r x + 1 - r)$ is monotone increasing in $r$, and therefore
\be\label{f-bd-1}
\sup_{x \ge 0} \sup_{\gamma \le r \le 1} \frac{f(x)}{f(rx + 1 -r)}
= \sup_{x \ge 0}  \frac{f(x)}{f(\gamma x + 1 - \gamma)}
\ee
The function $f(x) \, f(\gamma x + 1 - \gamma)^{-1}$ is analytic and decreasing at $x=1$
for $\gamma < 1$. 
Thus either the supremum in (\ref{f-bd-1})
is achieved at $x=0$ or else there is a critical point of the function
$f(x) \, f(\gamma x + 1 - \gamma)^{-1}$ in the interval $(0,\infty)$.
In order to rule out the second possibility,
we introduce a Lagrange multiplier and define the function
\be\label{def:Lag-mult}
h(x,y,\beta) = \log f(x) - \log f(y) - \beta (\gamma x + 1 - \gamma - y)
\ee
To find the critical points of $h$ we solve
\be
\frac{\partial h}{\partial x} = \frac{\partial h}{\partial y} = \frac{\partial h}{\partial \beta} = 0
\ee
Solving for $\beta$ leads to
\be
\frac{f'(x)}{f(x)} = \gamma \, \frac{f'(y)}{f(y)}
\ee
Since $y-1 = \gamma (x-1)$ this is equivalent to
\be\label{solve1}
\frac{(x-1) \log x}{x \log x - x + 1} = \frac{(y-1) \log y}{y \log y - y + 1}
\ee
Direct computation shows that
\be\label{deriv1}
\frac{d}{d x} \, \bigg( \frac{(x-1) \log x}{x \log x - x + 1} \bigg) & = &
f(x)^{-2} \, \Big((\log x)^2 - \frac{(x-1)^2}{x}\Big) \nonumber \\
& = & f(x)^{-2} \, (\log x)^2 \, \bigg(1 - \Big(\frac{x^{1/2}-x^{-1/2}}{\log x}\Big)^2 \bigg)
\ee
Furthermore, the function $x^{1/2}-x^{-1/2} - \log x$ is monotone increasing for all $x > 0$,
and thus $x^{1/2}-x^{-1/2} > \log x$ for $x > 1$.
Thus for $x \ge 1$ the derivative (\ref{deriv1}) is negative, and therefore
(\ref{solve1}) has no solution with $x > 1$.
Similarly $x^{1/2}-x^{-1/2} < \log x$ for $0 < x < 1$, and hence again
(\ref{deriv1}) is negative for $0 < x < 1$. So there are no solutions
of (\ref{solve1}) except $x=y=1$. Therefore (\ref{def:Lag-mult}) has no critical points
except $x=y=1$, and thus the function $f(x)f(\gamma x + 1 - \gamma)^{-1}$ achieves
its supremum at $x=0$.

\subsection{Proof of Lemma \ref{lem:M(h)}}\label{pf:lem:M(h)}
Suppose first that $0 < h < d \log d$. Recall the definition
\be\label{def:Mh2}
M_d(h) = \inf_{q \in {\Delta}_d} \bigg\{ \sum_{i=1}^d F(q_i d) \,:\, \sum_{i=1}^d f(q_i d) \ge h \bigg\}
\ee
where $F(x) = - \log x + x -1$ and $f(x) = x \log x - x + 1$. Letting $x_i = q_i d$ we have
\be\label{def:Mh3}
M_d(h) = \inf_{x_i \ge 0} \bigg\{ \sum_{i=1}^d F(x_i) \,:\, \sum_{i=1}^d f(x_i) \ge h, \qquad \sum_{i=1}^d x_i = d \bigg\}
\ee
The gradient of the function $\sum_{i=1}^d F(x_i)$ is zero only at $x_1 = \cdots = x_d = 1/d$,
hence since $h > 0$ there are no critical points of $\sum_{i=1}^d F(x_i)$ in the region
$\sum_{i=1}^d f(x_i) \ge h$. Thus the infimum in (\ref{def:Mh3}) is achieved at the boundary
where $\sum_{i=1}^d f(x_i) = h$, and so
\be\label{def:Mh4}
M_d(h) = \inf_{x_i \ge 0} \bigg\{ \sum_{i=1}^d F(x_i) \,:\, \sum_{i=1}^d f(x_i) = h, \qquad \sum_{i=1}^d x_i = d \bigg\}
\ee

\medskip
We introduce Lagrange multipliers and define
\be\label{def:Mh5}
H(x_i, \alpha, \beta) = \sum_{i=1}^d F(x_i) - \alpha \Big( \sum_{i=1}^d f(x_i) - h \Big) - \beta \Big(\sum_{i=1}^d x_i - d\Big)
\ee
The critical equations for $H$ are
\be
\frac{\partial H}{\partial x_i} = 1 - \frac{1}{x_i} - \alpha \log x_i - \beta = 0
\ee
The constraints can be used to eliminate $\beta$ and obtain
\be\label{def:Mh6}
\Big(1 + \alpha \frac{h}{d}\Big) x_i -1 = \alpha \, x_i \log x_i, \quad i=1,\dots,d
\ee
If $\alpha \le 0$ the equations (\ref{def:Mh6}) have the unique solution $x_i = 1$ for all
$i=1,\dots,d$. However this  does not satisfy the constraint $\sum_{i=1}^d f(x_i) = h$ for $h > 0$.
Thus $\alpha > 0$, in which case there are positive numbers $w$ and $z$ satisfying
\be
0 < w < 1 < z < d
\ee
such that the solutions of (\ref{def:Mh6}) are
\be
x_1 = \cdots = x_k = w, \quad x_{k+1} = \cdots = x_d = z
\ee
for some $1 \le k \le d-1$. The constraint conditions imply that
\be
k w + (d-k) z = d, \quad
k w \log w + (d-k) z \log z = h
\ee
Thus (\ref{def:Mh4}) can be reformulated as
\be\label{def:Mh7}
M_d(h) = \inf_{0 < w < z, \,\, 1 \le k \le d-1} \{ - k \log w - (d-k) \log z &:& k w \log w + (d-k) z \log z = h,\nonumber \\
&&  k w + (d-k) z = d \}
\ee
We claim that $- k \log w - (d-k) \log z$, subject to the constraints $k w \log w + (d-k) z \log z = h$ and
$k w + (d-k) z = d$, is a decreasing function of $k$. In order to show this, we divide (\ref{def:Mh7}) by $d$ and
write $k = t d$, and consider the function
\be\label{def:Mh8}
Q(w,z,t) = - t \log w - (1-t) \log z
\ee
along with the constraints
\be\label{def:Mh9}
t w \log w + (1-t) z \log z = \frac{h}{d}, \qquad t w + (1-t) z = 1
\ee
The constraints (\ref{def:Mh9}) allow $w,z$ to be defined locally as functions of $t$.
This follows from the implicit function theorem since the Jacobian is
$t(1-t) \log (w/z)$; we must have $w < z$ and hence the Jacobian is nonzero.
Solving these constraint equations for the derivatives gives
\be
\frac{d w}{d t} & = &  \frac{- w + z - w \log z + w \log w}{t \log (z/w)} \\
\frac{d z}{ d t} & = &  \frac{w - z - z \log w + z \log z}{(1-t) \log (z/w)}
\ee
Returning to (\ref{def:Mh8}) we can now compute its derivative with respect to $t$:
\be\label{def:Mh10}
\frac{d Q}{d t} & = & - \log w + \log z - \frac{t}{w} \frac{d w}{d t} - \frac{1-t}{z} \frac{d z}{ d t} \nonumber \\
& = & - \frac{1}{\log (z/w)} \bigg[ \frac{(z-w)^2}{zw} - \Big(\log (z/w)\Big)^2 \bigg]
\ee
Note that $2 \log u \le u - 1/u$ for all $u \ge 1$, hence
\be
\log \Big(\frac{z}{w}\Big) \le \sqrt{\frac{z}{w}} - \sqrt{\frac{w}{z}}
\ee
and therefore the right side of (\ref{def:Mh10}) is negative. Thus $Q$ is a decreasing function of
$t$, and hence the infimum in (\ref{def:Mh7}) is achieved at the largest possible value of $k$, namely $k=d-1$.
This leads to
\be\label{M-inf2}
M_d(h) = \inf_{z > 1} \{ - \log z - (d-1) \log \frac{d-z}{d-1} \,:\, z \log z + (d-z) \log \frac{d-z}{d-1} = h \}
\ee

\medskip
The function $z \log z + (d-z) \log \frac{d-z}{d-1}$ is monotone increasing, reaching its maximum value
$d \log d$ at $z=d$. Thus for any $0 < h < d \log d$ there is a unique value $z(d,h)$ satisfying the constraint condition in (\ref{M-inf2}). Its derivative is
\be\label{M-inf3}
\frac{\partial z}{\partial h} = \frac{d-z}{d \log z - h} \ge 0
\ee
Furthermore the function $g(z) = - \log z - (d-1) \log \frac{d-z}{d-1}$ is also monotone increasing for $1 < z < d$, with derivative $g'(z) = d (z-1)/z(d-z)$. Thus 
\be\label{M-inf4}
M_d(h) - M_d(0) & = & g(z(d,h)) - g(z(d,0)) \nonumber \\
& = & \int_{0}^h g'(z(d,h)) \, \frac{\partial z}{\partial h} \, d h \nonumber \\
& = & \int_{0}^h \frac{d(1 - z^{-1})}{d \log z - h} \, d h \nonumber \\
& \ge &  \int_{0}^h \frac{1}{z} \, \frac{d(z-1)}{d \log z - h} \, d h \nonumber \\
& \ge & \int_{0}^h \frac{1}{z} \, d h
\ee
The constraint condition in (\ref{M-inf2}) implies that
\be\label{M-inf5}
h \le z \log z \le h + (d-z) \, \log \frac{d-1}{d-z} \le h + z - 1
\ee
Thus
\be\label{M-inf6}
z \le \frac{h-1}{\log z -1}
\ee
If $h \ge 2 e^2$ then the first inequality in (\ref{M-inf5}) implies that
$z(d,h) \ge e^2$, and therefore $\log z \ge 2$. From (\ref{M-inf6}) it follows that
\be\label{M-inf7}
h \ge 2 e^2 \Rightarrow z(d,h) \le h -1
\ee
Thus from (\ref{M-inf4}) we deduce that for  $h \ge 2 e^2$
\be\label{M-inf7}
M_d(h) - M_d(0) \ge \int_{2 e^2}^h \frac{d h}{z} 
\ge \int_{2 e^2}^h \frac{d h}{h - 1} = \log (h-1) - \log (2 e^2 -1)
\ee
Since $z(d,0) = 1$ and $g(1)=0$ it follows that $M_d(0)=0$, and hence
(\ref{M-asymp}) holds.

\medskip
Finally, to show that $M_d(x)$ is increasing, note that
\be\label{M-inf8}
\frac{d M_d}{d x} = \Big[- \frac{1}{z} + \frac{d-1}{d-z} \Big] \, \frac{\partial z}{\partial x}
\ee
where $z$ solves the constraint equation
\be\label{M-inf9}
z \log z + (d-z) \log \Big(\frac{d-z}{d-1}\Big) = x
\ee
Differentiating (\ref{M-inf9}) gives
\be\label{M-inf10}
\frac{\partial z}{\partial x} = \Bigg( \log \big(\frac{z(d-1)}{d-z}\big) \Bigg)^{-1} > 0
\ee
since $z > 1$. Also $\Big[- \frac{1}{z} + \frac{d-1}{d-z} \Big] > 0$ hence
(\ref{M-inf8}) shows that $M_d$ is increasing.

\section{Discussion}
Hastings' Theorem finally settles the question of additivity of Holevo capacity for quantum channels,
as well as additivity of minimal output entropy and entanglement of formation.
In this paper we have explored in detail the proof of Hastings' result, and we have provided
some estimates for the minimal dimensions necessary in order to find a violation of
additivity. The violation of additivity seems to be a small effect for this class of models, requiring
delicate and explicit estimates for the proof. It is an open question whether there are random
unitary channels with large violations of additivity. Hastings' Theorem is non-constructive,
and it would be extremely interesting to find explicit channels which demonstrate the effect.
Presumably non-additivity of Holevo capacity is generic, and there may be other classes
of channels where the effect is larger.

\medskip
Having established non-additivity of Holevo capacity, one is led to the question of finding
useful bounds for the channel capacity $C(\Phi)$. One may even hope to find a 
compact `single-letter' formula for $C(\Phi)$, though that possibility seems remote.
It is likely that the methods introduced by Hastings will prove to be useful
in addressing these questions.

\medskip
\bigskip

\noindent{\bf Acknowledgments:}
M. F. thanks B. Nachtergaele, A. Pizzo and A. Soshnikov
for numerous discussions, 
M. Hastings for answering questions,
A. Holevo for useful comments on an early draft of this paper and
R. Siegmund-Schultze for sending his related slides. 
C. K. thanks P. Gacs, A. Harrow, T. Kemp, M. B. Ruskai, P. Shor and B. Zeng for 
useful conversations. This collaboration began
at the March 2009 workshop ``Entropy and the Quantum'' at the University of Arizona,
and the authors are grateful to the organizers of the workshop.

\pagebreak
\appendix

\section{Derivation of bound for $Z(n,d)$}
We consider the following integral.
\begin{align}
Z&=\int \delta \left(1-\sum_{i=1}^d p_i\right) \prod_{1\leq i< j\leq d } (p_i-p_j)^2 \prod_{k=1}^d p_k^{n-d}dp_k \\
&=\frac{1}{(dn-1)!}\int e^{-r}r^{dn-1}dr
\int \delta \left(1-\sum_{i=1}^d p_i\right) \prod_{1\leq i< j\leq d } (p_i-p_j)^2 \prod_{k=1}^d p_k^{n-d}dp_k 
\end{align}
Consider the following change of variables.
\begin{align}
q_1&=rp_1\\
\vdots \\
q_{d-1}&=rp_{d-1}\\
q_d&=r(1-p_1-\ldots -p_{d-1})
\end{align}
The Jacobian is
\begin{align}
\frac{\partial (q_1, \ldots, q_d)}{\partial(p_1, \ldots, p_{d-1},r)}
&=\begin{vmatrix}
r & \ldots & 0&p_1 \\
\vdots& \ddots & \vdots & \vdots \\
0 &\ldots & r & p_{d-1} \\
-r&\ldots & -r & 1-p_1-\ldots -p_{d-1}
\end{vmatrix}
=\begin{vmatrix}
r & \ldots & 0&p_1 \\
\vdots& \ddots & \vdots & \vdots \\
0 &\ldots & r & p_{d-1} \\
0&\ldots & 0 & 1
\end{vmatrix}
=r^{d-1}
\end{align}
After the change of variables we have
\begin{align}
Z=\frac{1}{(dn-1)!}
\int \prod_{1\leq i< j\leq d } (q_i-q_j)^2 \prod_{k=1}^d e^{-q_k}q_k^{n-d}dq_k.
\end{align}
However,
\begin{align}
\prod_{1\leq i< j\leq d } (q_i-q_j)^2 
=\begin{vmatrix}
1 & \ldots & 1\\
q_1 & \ldots & q_d\\
\vdots &\ddots &\vdots \\
q_1^{d-1} & \ldots & q_d^{d-1}\\
\end{vmatrix}^2
=\begin{vmatrix}
1 & \ldots & 1\\
p_1(q_1) & \ldots & p_1(q_d)\\
\vdots &\ddots &\vdots \\
p_{d-1}(q_1) & \ldots & p_{d-1}(q_d)
\end{vmatrix}^2.
\end{align}
Here, $p_k$ is any monic polynomial of degree $k$.
So, set $p_k = (-1)^kk!L_k^{n-d}$, where $L_k^{n-d}$ is the Laguerre polynomial.
Then, we have
\begin{align}
\int p_k (x) p_l(x)\;e^{-x}x^{n-d} dx = \delta_{kl}\Gamma(n-d+k+1)\Gamma(k+1)
\end{align}
Hence
\begin{align}
Z &=\frac{1}{(dn-1)!}
\int \left( \sum_\sigma {\rm sign}(\sigma) \prod_{i=1}^d p_{\sigma(i)}(q_i) \right)^2 \prod_{k=1}^d e^{-q_k}q_k^{n-d}dq_k\\
&= \frac{1}{(dn-1)!}
\sum_\sigma \prod_{k=1}^d \int    \left( p_{\sigma(k)}(q_k)\right)^2  e^{-q_k}q_k^{n-d}dq_k\\
&=\frac{\Gamma(d+1)}{\Gamma(dn)}
\prod_{k=0}^{d-1} \Gamma( n-d +k+1) \Gamma(k+1)\\
&= \frac{1}{\Gamma(dn)}
\prod_{k=1}^{d} \Gamma( n-d +k) \Gamma(k+1).
\label{integral}
\end{align}
Here, $\sigma$ are ``permutations''; $\sigma: \{1,\ldots,d \} \rightarrow \{0,\ldots,d-1 \}$.

To evaluate this quantity we use the following fact:
$\Gamma (s)$ is approximated by
\begin{align}
\exp \left\{ s\ln s -s - \frac{1}{2}\ln s  
+ \ln \sqrt{2\pi} + \ln \left( 1+ O\left( \frac{1}{|s|^{\frac{1}{2}}}\right)\right)  \right\}
\label{gammaapprox}  
\end{align}
as $s\rightarrow +\infty$. Then, we have
\begin{align}
\exp\{(s-1)\ln s -s \} < (\ref{gammaapprox}) < \exp\{s\ln s -s\}.
\end{align}
Note that the above upper bound is true only for large enough $s$ but
in our case it is not a problem.
By using these bounds we get a lower bound for (\ref{integral}).

First,
$\ln(1/ \Gamma(dn)))$ is lower bounded by
\begin{align} 
 - (dn)\ln(dn) +dn = -dn \ln d -dn\ln n +dn. 
\label{app1}
\end{align} 
Secondly,
$\ln(\prod_{k=1}^d \Gamma( n-d +k))$ is lower bounded by
\begin{align} 
&\sum_{k=1}^d (n-d +k-1) \ln (n-d +k) 
- \sum_{k=1}^d (n-d +k)\\
&= n^2  \sum_{k=1}^d \frac{1}{n}\frac{n-d +k-1}{n} \ln \left(\frac{n-d +k}{n}\right)
+ \sum_{k=1}^d (n-d +k-1) \ln n 
- \sum_{k=1}^d (n-d +k). 
\label{app2}
\end{align}
The first sum in(\ref{app2}) is approximately lower bounded by
\begin{align}
n^2 \times \frac{d}{n} \times \frac{n-d }{n} \ln \left(\frac{n-d}{n}\right)
= d(n-d)(\ln (n-d) - \ln n)\approx -d^2,
\end{align}
as $n\rightarrow \infty$.
Also, the remaining  part in (\ref{app2}) is
\begin{align}
\left(dn- \frac{1}{2}d^2 - \frac{1}{2}d \right)\ln n -\left(dn- \frac{1}{2}d^2 + \frac{1}{2}d \right).
\label{app3} 
\end{align}
Thirdly, 
$\ln (\prod_{k=1}^d  \Gamma(k+1))$ is lower bounded by 
\begin{align}
\sum_{k=1}^d & \;k \ln (k+1) -\sum_{k=1}^d (k+1) \\
&= d^2  \sum_{k=1}^d \frac{1}{d}\frac{k}{d} \ln \left(\frac{k+1}{d}\right)
+ \sum_{k=1}^d \;k \ln d -\sum_{k=1}^d (k+1) .\label{app4}
\end{align}
Again,
the first term in (\ref{app4}) is approximately lower bounded by $-d^2$.
Also, 
the remaining part in (\ref{app4}) is 
\begin{align}
\left(  \frac{1}{2}d^2 +\frac{1}{2}d \right)\ln d -\left(  \frac{1}{2}d^2 +\frac{3}{2}d \right).
\label{app5}
\end{align}
As a whole, we know that
the inside of $\exp$ in (\ref{integral}) is lower bounded by
\begin{align}
(\ref{app1})&+ (\ref{app3})+ (\ref{app5})-2d^2 \\
&=\left[ -\frac{1}{2}d^2 - \frac{1}{2}d \right]\ln n 
+ \left[-dn +\frac{1}{2}d^2 + \frac{1}{2}d \right]\ln d 
-2d^2- 2d \\
&= -d^2\ln n + (d^2-dn) \ln d 
+\frac{1}{2} d(d-1)(\ln n - \ln d -4) \\
&  \geq -d^2\ln n + (d^2-dn) \ln d
\end{align}
if $(\ln n - \ln d) \geq 4$.
Therefore, we get an upper bound for the normalization constant:
\begin{align}
Z(n,d)^{-1} \le 
 n^{d^2} d^{d(n-d)}
\end{align}
in this case.

\section{Proof of Proposition \ref{random-states}}
Let $Z_1,\dots,Z_n$ be IID complex Gaussian
random vectors with mean zero and variance one. Apply the result (\ref{lem1pf:12}) with
$m=n-1$ to deduce that
\be
(Z_2,\dots,Z_n)^T = \rho \, | \phi \ket
\ee
where $| \phi \ket$ is a random unit vector in 
$ \mathbb{C}^{n-1} \hookrightarrow \mathbb{C}^{n} $, independent of 
$\rho = (|Z_2|^2+ \cdots + |Z_{n}|^2)^{1/2}$. Then apply (\ref{lem1pf:12}) with $m=n$ to deduce
\be
(Z_1,\dots,Z_n)^T = R \, | \theta \ket
\ee
where $|\theta\ket$ is a random unit vector in $\mathbb{C}^{n}$, independent of 
$R = (|Z_1|^2+ \cdots + |Z_{n}|^2)^{1/2}$. Let
\be\label{x-frac1}
x = \frac{Z_1}{R} = \frac{Z_1}{\sqrt{|Z_1|^2 + \rho^2}}
\ee
and recall that $| \psi \ket = (1,0,\dots,0)^T$. Then
\be
| \theta \ket = \frac{1}{R} \, (Z_1,\dots,Z_n)^T = x | \psi \ket + \frac{\rho}{R} \, | \phi \ket = x | \psi \ket + \sqrt{1 - |x|^2} \, | \phi \ket
\ee
This proves the first part Proposition \ref{random-states} since $|\phi \ket$ is independent of $Z_1$ and $\rho$, and hence is independent of $x$. 

\medskip
To prove the second part of Proposition  \ref{random-states} we use the representation
(\ref{x-frac1}) to derive the distribution of $|x|^2$. We write $Z_j = X_{2j-1} + i X_{2j}$ where
$\{X_j\}$ are IID real normal random variables with mean zero and variance one, so from
(\ref{x-frac1}) it follows that
\begin{align}
|x|^2 = \frac{X_1^2 + X_2^2}{X_1^2 + \ldots + X_{2n}^2}.
\end{align}
Note that $X_1^2 + \ldots + X_k^2$ has 
the following Chi-square probability distribution:
\begin{align}
f_k (x) =
\frac{1}{2^{\frac{k}{2}}\Gamma(\frac{k}{2})} x^{\frac{k}{2}-1}e^{-\frac{x}{2}}
\end{align}
Hence, set
\begin{align}
X&= X_1^2 + X_2^2 \\
Y&= X_3^2 + \ldots + X_{2n}^2
\end{align}
and then $X$ and $Y$ are independent and have the following probability distributions
\begin{align}
f_X (x) &= \frac{1}{2}e^{-\frac{x}{2}} \\
f_Y (y) 
&=
\frac{1}{2^{n-1}\Gamma(n-1)} y^{n-2}e^{-\frac{y}{2}}
\end{align}
However,
\begin{align}
\frac{X}{X+Y} \leq t \Leftrightarrow X(1-t) \leq tY 
\end{align}
implies that the cumulative function of $\frac{X}{X+Y}$ is
\begin{align}
\int_0^\infty \left(\int_0^{\frac{t}{1-t}y}  f_X(x)\; dx \right) f_Y (y) \; dy 
&= \int_0^\infty \left( 1- e^{-\frac{ty}{2(1-t)}} \right)f_Y (y) \; dy \\
&= 1- \int_0^\infty  e^{-\frac{ty}{2(1-t)}} f_Y (y) \; dy \\
&=1- \frac{1}{2^{n-1} (n-2)!}\int_0^\infty y^{n-2}e^{-\frac{y}{2(1-t)} }
\end{align}
Here,
\begin{align}
 \int_0^\infty y^{n-2}e^{-\frac{y}{2(1-t)} } 
&= \left( 2(1-t) \right)^{(n-2)} (n-2)! \int_0^\infty e^{-\frac{y}{2(1-t)} } \; dy \\
&= 2^{n-1} (1-t)^{n-1} (n-2)!.
\end{align}
Therefore
\begin{align}
F_{\frac{X}{X+Y}}(t)= 1- (1-t)^{n-1}.
\end{align}

\end{document}